\documentclass[traditabstract,epsf]{aa}
\usepackage{txfonts}
\usepackage{epsfig}

\newcommand{\teff}{\mbox{$T_{\rm eff}$}}
\newcommand{\logg}{\mbox{$\log g$ }}
\newcommand{\vsini}{\mbox{$v \sin{i_\ast}$ }}
\newcommand{\mictrb}{\mbox{$\xi_{\rm t}$}}
\newcommand{\mactrb}{\mbox{$v_{\rm mac}$}}

\newcommand{\kms}{\mbox{km\,s$^{-1}$ }}
\newcommand{\ms}{\mbox{m\,s$^{-1}$ }}
\newcommand{\halpha}{\mbox{$H_\alpha$}}

\begin{document}
\title{WASP-64\,b and WASP-72\,b: two new transiting highly irradiated giant planets\thanks{The photometric time-series used in this work are only available in electronic form at the CDS via anonymous ftp to  cdsarc.u-strasbg.fr (130.79.128.5) or via http://cdsweb.u-strasbg.fr/cgi-bin/qcat?J/A+A/}}

\author{ 
M.~Gillon\inst{1}, D.~R.~Anderson\inst{2}, A.~Collier-Cameron \inst{3},  A.~P.~Doyle\inst{2}, A.~Fumel\inst{1}, C.~Hellier\inst{2}, E.~Jehin\inst{1},
M.~Lendl\inst{4}, P.~F.~L.~Maxted\inst{2}, J.~Montalb\'an\inst{1}, F.~Pepe\inst{4}, D.~Pollacco\inst{5}, D.~Queloz\inst{4}, D.~S\'egransan\inst{4},
A.~M.~S.~Smith\inst{2}, B.~Smalley\inst{2}, J.~Southworth\inst{2}, A.~H.~M.~J.~Triaud\inst{4}, S.~Udry\inst{4},  R.~G. West\inst{6}}

\offprints{michael.gillon@ulg.ac.be}
\institute{
    $^1$ Institut d'Astrophysique et de G\'eophysique, Universit\'e de Li\`ege, All\'ee du 6 ao\^ut 17, Sart Tilman, Li\`ege 1, Belgium \\
    $^2$ Astrophysics Group, Keele University, Staffordshire, ST5 5BG, United Kingdom\\
    $^3$ School of Physics and Astronomy, University of St. Andrews, North Haugh, Fife, KY16 9SS, UK\\
    $^4$ Observatoire de Gen\`eve, Universit\'e de Gen\`eve, 51 Chemin des Maillettes, 1290 Sauverny, Switzerland\\  
    $^5$ Department of Physics, University of Warwick, Coventry CV4 7AL, UK\\
    $^6$ Department of Physics and Astronomy, University of Leicester, Leicester, LE1 7RH, UK\\
    }   

\date{Received date / accepted date}
\authorrunning{M. Gillon et al.}
\titlerunning{WASP-64\,b and WASP-72\,b}
\abstract{
We report the discovery by the WASP transit survey of two new highly irradiated giant planets.
 WASP-64\,b is slightly more massive ($1.271 \pm 0.068$ $M_{Jup}$) and larger ($1.271 \pm 
 0.039$ $R_{Jup}$) than Jupiter, and is in very-short ($a=0.02648 \pm 0.00024$ AU, 
$P=1.5732918 \pm 0.0000015$  days) circular orbit around a $V$=12.3 
 G7-type dwarf ($1.004 \pm 0.028$ $M_\odot$, $1.058 \pm 0.025$ $R_\odot$, 
\teff   = 5500 $\pm$ 150 K). Its size is typical of hot Jupiters with similar masses. 
WASP-72\,b has also a mass a bit higher than Jupiter's ($1.461_{-0.056}^{+0.059}$
$M_{Jup}$) and orbits very close ($0.03708 \pm 0.00050$ AU, 
$P=2.2167421 \pm 0.0000081$  days) to a bright ($V$=9.6) and moderately evolved 
 F7-type star ($1.386 \pm 0.055$ $M_\odot$, $1.98 \pm 0.24$ $R_\odot$, 
\teff   = 6250 $\pm$ 100 K).  Despite its extreme irradiation ($\sim5.5\times10^9$ erg\,s$^{-1}$\,cm$^{-2}$), 
WASP-72\,b has a moderate size ($1.27 \pm 0.20$ $R_{Jup}$) that could suggest a significant 
enrichment in heavy elements.  Nevertheless, the errors
on its physical parameters  are still too high to draw any strong inference on 
its internal structure or its possible peculiarity.
}

\maketitle

\section{Introduction}
The booming study of exoplanets allow us to assess the diversity of the planetary 
systems of the Milky Way and to put our own solar system in perspective. Notably, 
ground-based  transit surveys targeting relatively bright ($V<13$) stars are detecting 
at an increasing rate  short-period giant planets amenable for a thorough characterization
(orbit, structure, atmosphere), thanks to the brightness of their host star, the favorable 
planet-star size ratio and their high stellar irradiation (e.g. Winn 2010).  With its very high detection
efficiency, the WASP transit survey (Pollacco et al. 2006) is one of the most productive 
projects in that domain.

In this context, we report here the detection by WASP of two new giant planets,
 WASP-64\,b and WASP-72\,b, transiting relatively bright Southern stars.
 Section~2 presents the WASP discovery photometry, and  high-precision 
 follow-up observations  obtained from La Silla ESO Observatory (Chile) by the TRAPPIST and 
 {\it Euler} telescopes to confirm the transits and planetary nature of both objects and to 
 determine precisely the systems parameters. In Sect.~3, we present the spectroscopic 
 determination of the stellar properties and the derivation of the systems parameters  through 
 a combined analysis of the follow-up photometric and spectroscopic time-series.
Finally, we discuss our results in Sect.~4.

\section{Observations}

\subsection{WASP transit detection photometry}

The stars 1SWASPJ064427.63-325130.4 (WASP-64; $V$=12.3, $K$=11.0) 
and 1SWASPJ024409.60-301008.5 (WASP-72; $V$=10.1, $K$=9.6)
 were observed by the Southern station of the WASP survey (Hellier et al. 2011) 
 between 2006 Oct 11 and 2010 Mar 12 and between 2006 Aug 11 and 2007 
 Dec 31, respectively. The 17981 and 6500 pipeline-processed photometric 
 measurements were detrended and  searched for transits using the  methods 
  described by Collier-Cameron et al. (2006). The selection process 
  (Collier-Cameron et al. 2007) identified WASP-72 as a  high priority 
  candidate  showing periodic low-amplitude (2-3 mmag) transit-like signatures
   with period of  2.217~days. For WASP-64, similar transit-like signals with a period of 
   1.573~days were also detected,  not only on the target itself but also on a brighter star at 28'', 
   1SWASPJ064429.53--325129.5 (TYC7091-1288-1, $V$=12.3, $K$=11.0).   Fig.~1 presents for 
TYC7091-1288-1 and WASP-64  the WASP photometry folded on the deduced transit ephemeris. 
Fig.~2 does the same for WASP-72.

A search for periodic modulation was  applied to the photometry of 
WASP-72, using for this purpose the method described in
 Maxted et al. (2011). No periodic signal was found down to the mmag amplitude.
  We did not perform such a  search for TYC7091-1288-1 and WASP-64, as these two
 stars  are blended together at the spatial resolution of the WASP instrument (see below).
 Still, a Lomb-Scargle periodogram analysis of their photometric time-series did
 not reveal any significant power excess.

\subsection{Follow-up photometry}

\subsubsection{WASP-64}

WASP-64 is at 28'' West from TYC7091-1288-1, close
 enough to have most of its point-spread function (PSF) enclosed in the smallest
 of the WASP photometry extraction apertures (radius=34'', see Fig.~3). 
Both objects have an entry in the WASP database,  because it is based 
 on an input catalogue of star positions. Still, the WASP light curve obtained with an 
 aperture centered on WASP-64 is of poorer quality (see Fig.~1), because the centering algorithm 
 does not work optimally  when there is a bright object off-centre in the aperture or
  just outside of it,  while significant levels of red noise are brought by PSF variations. This 
  explains why the transit was first detected from the photometry centered on TYC7091-1288-1.
Having both stars nearly totally enclosed in the smallest apertures for both centerings prevented us to decide
from the WASP photometry alone  if the eclipse signal detected by WASP was originating  from
 one or the other star, so our first follow-up action was to measure on 2011 Jan 20 a transit at a 
 better spatial resolution with the robotic 60cm telescope TRAPPIST ({\it TRA}nsiting
{\it P}lanets and {\it P}lanetes{\it I}mals {\it S}mall {\it T}elescope; Gillon et al. 2011, 
Jehin et al. 2011) located at ESO La Silla Observatory in the Atacama Desert, Chile. 
TRAPPIST is equipped with a thermoelectrically-cooled 2K\,$\times$\,2K CCD having
 a pixel scale of 0.65'' that translates into a 22'\,$\times$\,22'  field of view. Differential 
 photometry was obtained with TRAPPIST for both stars on the night of 2011 Jan 20, 
 corresponding to a transit window as derived from WASP data. These observations 
 were obtained with the telescope focused and through a special `$I+z$' filter that has a
transmittance $>$90\% from 750 nm to beyond 
1100 nm\footnote{http://www.astrodon.com/products/filters/near-infrared/}.  
The positions of the stars on the chip were maintained to within a few pixels over the course of the run, 
thanks to a `software guiding' system deriving 
regularly  an astrometric solution for the most recently acquired image and sending
 pointing corrections to the mount if needed.  After a standard pre-reduction (bias, dark, 
 flatfield correction), the stellar fluxes were extracted from the images using the {\tt IRAF/DAOPHOT}\footnote{{\tt IRAF} is 
 distributed by the National Optical Astronomy Observatory, which is 
 operated by the Association of Universities for Research in Astronomy, Inc., under cooperative agreement
  with the National Science Foundation.} aperture photometry software (Stetson, 
 1987). Several sets of reduction parameters were tested, and we kept the one giving the 
 most precise photometry for the stars of similar brightness as the target. After a careful
  selection of a set of 22 reference stars, differential photometry was then obtained.  This reduction 
  procedure was also applied for the subsequent TRAPPIST runs.

 This first TRAPPIST run resulted in a flat light curve for TYC7091-1288-1, 
 while the light curve for WASP-64 showed a clear  transit-like structure (Fig.~3), identifying thus  
WASP-64 as the source of the transit signal. A second (partial) transit was observed in the 
 $I+z$ filter on 2011 Feb 22 to better constrain  the shape of the eclipse (Fig.~4, second light
  curve from the top). As for the following WASP-64 transits, the telescope was defocused
  to $\sim$3'' to improve the duty cycle and average the pixel-to-pixel effects. A global analysis 
  of the two first TRAPPIST  transit light curves led to an eclipse depth and shape compatible 
 with the transit of a giant planet in front of a solar-type star.  Our next action was to observe 
 a third transit with TRAPPIST, this time in the $V$ filter to assess the chromaticity of the 
 transit depth (Fig.~4, third light curve from the top). The analysis of the resulting light curve 
 led to a transit depth consistent with the one measured in the $I+z$ filter, as expected for a 
 transiting planet. We then observed an $occultation$ window in the $z'$-band on 2011 Apr 30. 
 We could not detect any eclipse in the resulting photometric time-series (Fig.~5), which was again
  consistent with the transiting planet scenario. At this stage, we began our spectroscopic 
 follow-up of WASP-64 that confirmed the solar-type nature of WASP-64 and the planetary nature
  of its eclipsing companion (see Sec. 2.3).

Once the planetary nature of WASP-64\,b was confirmed, we observed seven more of its transits with 
TRAPPIST, using then a blue-blocking filter\footnote{http://www.astrodon.com/products/filters/exoplanet/} 
that has a transmittance  $>$90\% from 500 nm to beyond 1000 nm.  The goal of using this very 
wide red filter is to maximize the signal-to-noise ratio (S/N) while minimizing the influence of
 moonlight pollution, differential extinction and stellar limb-darkening on the transit light curves. 
 The resulting light curves are shown in Fig.~4.
The transit of 2011 Oct 19 was also observed in the Gunn-$r$ filter with the EulerCam CCD camera 
at the 1.2-m {\it Euler} Telescope at La Silla Observatory. This nitrogen-cooled camera has a 
4k\,$\times$\,4k E2V CCD with a 15'\,$\times$\,15'  field of view  (scale=0.23"/pixel). Here too, 
a defocus was applied to the telescope to optimize the observation efficiency and minimize 
pixel-to-pixel effects, while flat-field effects were further reduced by keeping the stars on the same 
pixels, thanks to a `software guiding' system similar to the one of TRAPPIST (Lendl et al. 2012). 
The reduction was similar to that performed on TRAPPIST data. The resulting light curve is also 
shown in Fig.~4.
 
\subsubsection{WASP-72}

We monitored for WASP-72 five transits with TRAPPIST (see Table 1 and Fig.~6),  two partial and
 two full transits in the $I+z$ filter and one full transit in the blue-blocking filter. For the three transits
  observed in 2011, the telescope was defocused to $\sim$3''.  A first partial transit was observed in 
  2011 Jan 21 in the $I+z$ filter at high airmass, confirming the low-amplitude eclipse detected by  
  WASP (Fig.~6, first light curve from the top). The next season, a full transit was observed on 2011 
  Oct 25 in the blue-blocking filter. A technical problem damaged these data: a shutter  problem led to 
  a scatter twice higher than expected.  In 2011,  a partial transit was also observed in the  $I+z$ filter on 
  Dec 4.  In 2012, two new full transits were observed with TRAPPIST in the $I+z$ filter. For these two last 
  runs, the telescope was kept focused to minimize the effects of a  focus drift problem with an amplitude 
  stronger for out-of-focus observations. We are still investigating the origin of this technical problem. 
  Two transits of WASP-72 were also observed with {\it Euler} on 2011 Nov 26 and 2012 Nov 16,  with the 
  same strategy than for WASP-64. For the second {\it Euler} transit, a crash of the tracking system led to significant
  shifts of the stars on the detectors (up to 50 pixels), giving rise to significant systematic effects in the differential 
  photometry (see Fig.~6)
\\
\\
Table 1 presents a summary of the follow-up photometric time-series obtained for WASP-64 and WASP-72.  
 
\begin{figure}
\label{fig:wasp64+TYC}
\centering                     
\includegraphics[width=9cm]{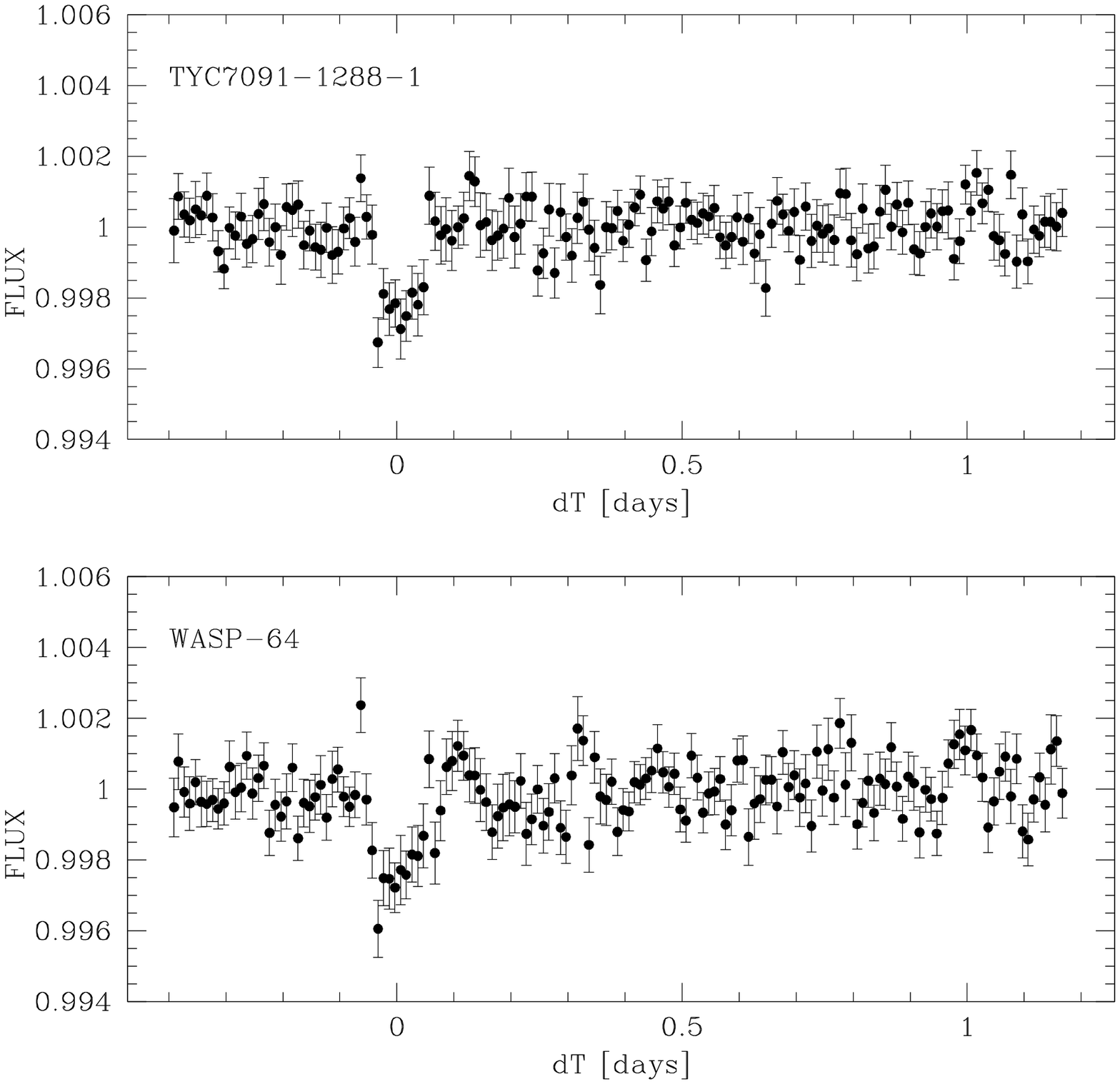}
\caption{WASP photometry for TYC7091-1288-1 ($top$) and  WASP-64 ($bottom$) 
folded on the best-fitting transit ephemeris from the transit search algorithm presented in Collier Cameron et al. (2006), 
and binned per 0.01d intervals.}
\end{figure}

\begin{figure}
\label{fig:wasp72}
\centering                     
\includegraphics[width=9cm]{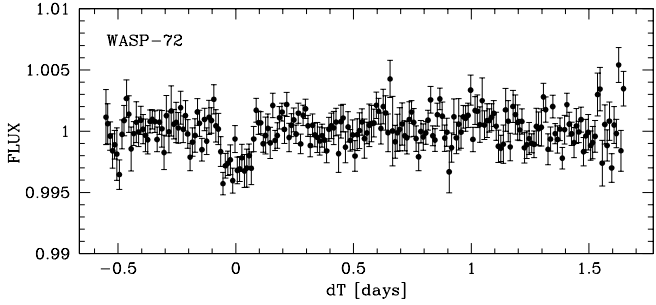}
\caption{WASP photometry for WASP-72 
folded on the best-fitting transit ephemeris from the transit search algorithm presented in Collier Cameron et al. (2006), 
and binned per 0.01d intervals.}
\end{figure}

\begin{figure*}
\label{fig:w64ima}
\centering                     
\includegraphics[width=18cm]{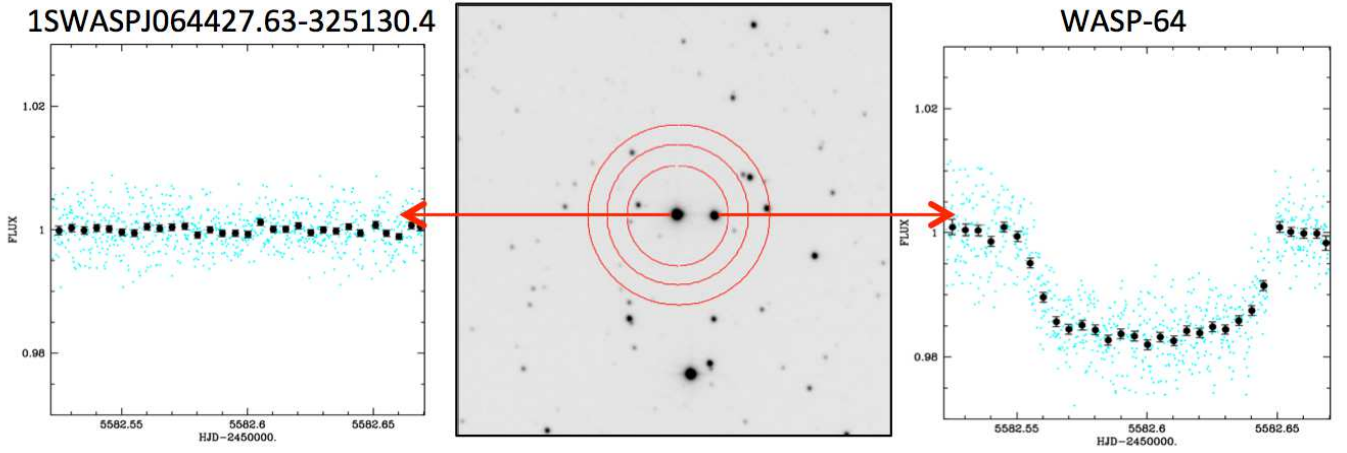}
\caption{280''$\times$280'' TRAPPIST $I+z$ image centered on TYC7091-1288-1. North is up and East is left.  
The three concentric circles indicate the three photometry extraction apertures used in the WASP pipeline. 
WASP-64 is  the closest star to the right of TYC7091-1288-1. For both stars, the light curve
obtained by TRAPPIST on 2011 Jan 20 is shown (cyan=unbinned, black=binned per intervals of
0.005d).
}
\end{figure*}

\begin{figure}
\label{fig:w64phot}
\centering                     
\includegraphics[width=9cm]{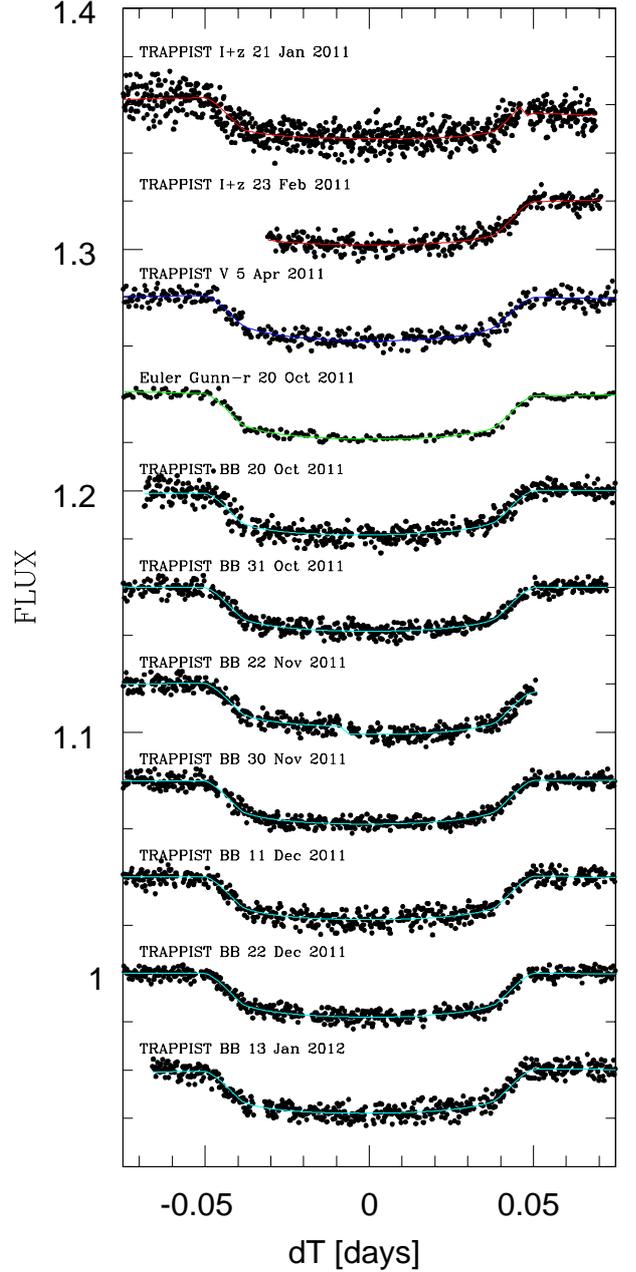}
\caption{Follow-up transit photometry for WASP-64\,b.  For 
each light curve, the best-fit transit+baseline model deduced from the global analysis is superimposed
(see Sec. 3.2). The light curves are shifted along the $y$-axis for clarity.  BB = Blue-blocking filter.}
\end{figure}

\begin{figure}
\label{fig:w64occ}
\centering                     
\includegraphics[width=8.5cm]{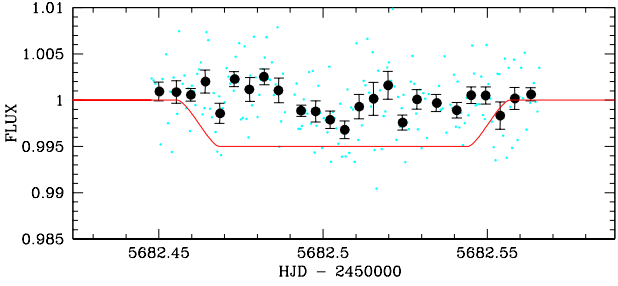}
\caption{TRAPPIST $z'$ time-series photometry obtained during an occultation window of WASP-64\,b, unbinned and 
binned per intervals of 0.005d. An occultation model assuming a circular orbit and a depth of 0.5\% is superimposed
for comparison.  }
\end{figure}

\begin{figure}
\label{fig:w72phot}
\centering                     
\includegraphics[width=9cm]{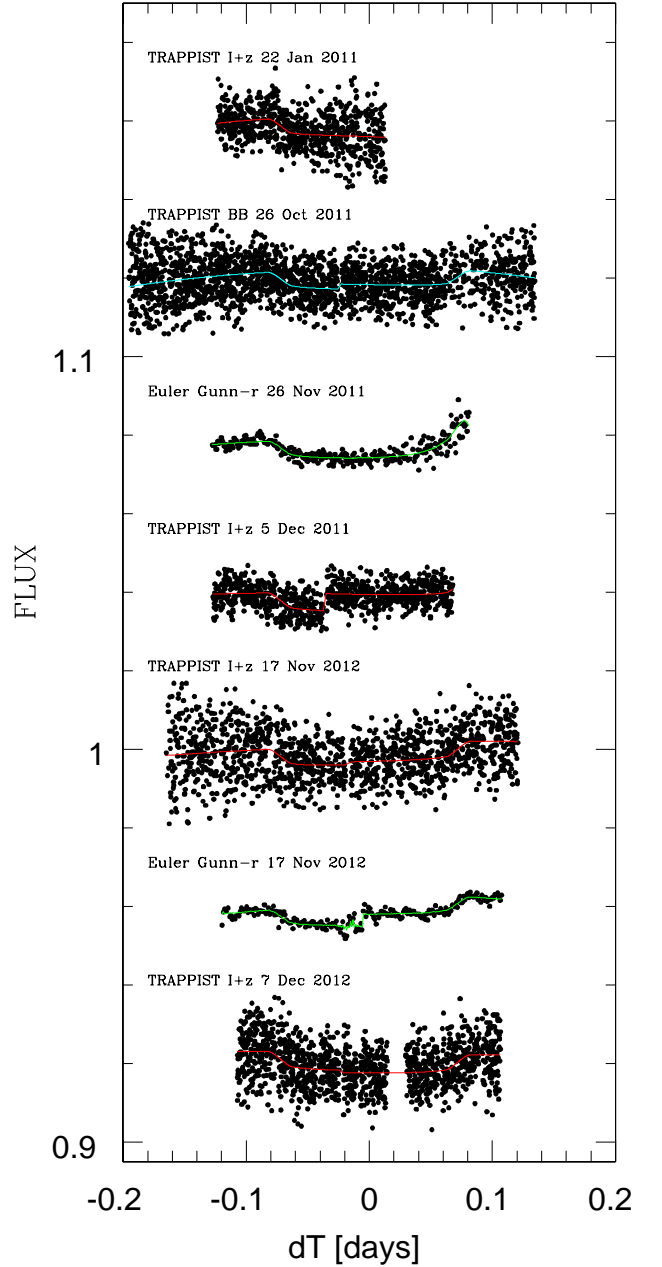}
\caption{Follow-up transit photometry for WASP-72\,b. For each light curve, the best-fit
 transit+baseline model deduced from the global analysis is superimposed (see Sec. 3.2). The light curves 
 are shifted along the $y$-axis for clarity.}
\end{figure}

\begin{table*}
\begin{center}
\begin{tabular}{cccccccc}
\hline
Target &      Night    & Telescope & Filter & $N$         & $T_{exp}$ &  Baseline function  & Eclipse\\ 
             &                   &                    &            &                 &       (s)           &                                  &  nature              \\ \hline \noalign {\smallskip} 
WASP-64  &  2011 Jan 20-21 & TRAPPIST &  $I+z$ & 792   & 8  &  $p(t^2)+o$  & transit               \\ \noalign {\smallskip} 
WASP-64  &  2011 Feb 22-23  & TRAPPIST &  $I+z$ & 296  & 20 &  $p(t^2)$   & transit                \\ \noalign {\smallskip} 
WASP-64   &  2011 Apr 4-5    &  TRAPPIST   &  $V$ & 364   & 30   &  $p(t^2)$    &  transit   \\ \noalign {\smallskip} 
WASP-64   &  2011 Apr 30 - May 1& TRAPPIST &     $z'$ & 202    &  40 & $p(t^2)$ &   occultation              \\ \noalign {\smallskip} 
WASP-64  & 2011 Oct 19-20   & TRAPPIST & BB &   578 &  15  &  $p(t^2)$  &   transit                \\ \noalign {\smallskip} 
WASP-64  & 2011 Oct 19-20   & {\it Euler} & Gunn-$r$ &   159 &  60  &  $p(t^2)$    &  transit                \\ \noalign {\smallskip} 
WASP-64  & 2011 Oct 30-31   & TRAPPIST & BB  &    533    &     15     &        $p(t^2)$     &  transit        \\ \noalign {\smallskip} 
WASP-64  & 2011 Nov 21-22   & TRAPPIST & BB  &   512     &   15   &  $p(t^2)+o$     & transit                  \\ \noalign {\smallskip} 
WASP-64  & 2011 Nov 29-30   & TRAPPIST & BB  &    611    &    15          &        $p(t^2)$ &  transit        \\ \noalign {\smallskip} 
WASP-64  & 2011 Dec 10-11   & TRAPPIST & BB  &  642    &   15    &    $p(t^2)+o$   & transit       \\ \noalign {\smallskip} 
WASP-64 & 2011 Dec 21-22   & TRAPPIST & BB  &   566     & 15    &    $p(t^2)+o$ & transit   \\ \noalign {\smallskip} 
WASP-64 & 2012 Jan 12-13   & TRAPPIST & BB  & 578       &     15    &        $p(t^2)$ &  transit        \\[0.33cm] 
WASP-72 & 2011 Jan 21-22   & TRAPPIST & $I+z$  &  707      &        8          &     $p(t^2)$         & transit   \\ \noalign {\smallskip} 
WASP-72 & 2011 Oct 25-26   & TRAPPIST & BB  &  2042      &   4     &  $p(t^2)+o$   & transit   \\ \noalign {\smallskip} 
WASP-72 & 2011 Nov 25-26 & {\it Euler} & Gunn-$r$ & 294 & 40 & $p(t^2)+p(a^2)$ & transit \\ \noalign {\smallskip} 
WASP-72 & 2011 Dec 4-5   & TRAPPIST & $I+z$ &  892      &       10           &   $p(t^2)+o$         & transit    \\ \noalign {\smallskip} 
WASP-72 & 2012 Nov 16-17   & TRAPPIST & $I+z$ &  1344      &       6          &   $p(t^2)+o$         & transit    \\ \noalign {\smallskip} 
WASP-72 & 2012 Nov 16-17   & {\it Euler} & Gunn-$r$ &  217     &       70          &   $p(t^2+xy^1)+o$      & transit    \\ \noalign {\smallskip} 
WASP-72 & 2012 Dec 6-7   & TRAPPIST & $I+z$ &  1197     &       6           &   $p(t^2)+o$         & transit    \\ \noalign {\smallskip} 
\hline
\end{tabular}
\caption{Summary of follow-up photometry obtained for WASP-64 and WASP-72. $N$=
number of measurements. $T_{exp}$ = exposure time. BB = blue-blocking
filter. The baseline functions are the analytical functions used to model the photometric baseline of each
light curve (see Sec. 3.2). $p(t^2)$ denotes a quadratic time polynomial, $p(a^2)$ a quadratic
airmass polynomial, $p(xy^1)$ a linear function of the stellar position on the detector, and $o$ an offset fixed at the time of 
the meridian flip.}
\end{center}
\label{wasp64+72_trappist}
\end{table*}

\subsection{Spectroscopy and radial velocities}

Once WASP-64 and WASP-72 were identified as  high priority candidates, we gathered spectroscopic measurements 
with the {\tt CORALIE} spectrograph mounted on {\it Euler} to confirm the planetary nature of the eclipsing bodies and 
obtain mass measurements. 16 usable spectra were obtained for WASP-64 from 2011 May 2 to 2011 November 7 
with an exposure time of 30 minutes. For WASP-72, 18 spectra were gathered from 2011 January 9 to 2011 December 29, 
here too with an exposure time of 30 minutes.  For both stars, radial velocities  (RVs)  were computed by weighted cross-correlation
 (Baranne et al. 1996) with a numerical G2-spectral template giving close to optimal precisions for late-F to early-K dwarfs, from our 
 experience.  The resulting RVs are shown in Table~2. 

The RV time-series show variations that are  consistent with planetary-mass companions. Preliminary orbital analyses 
of the RVs resulted in periods and phases in excellent agreement with those deduced from the WASP transit detections (Fig.~7 \& 8, 
upper panels).  For WASP-64, assuming a stellar mass $M_{\ast} = 0.98 \pm 0.09$ $M_{\sun}$  (Sect.~3.1), the fitted semi-amplitude 
$K = 212 \pm 17$ m\,s$^{-1}$ translates into a secondary mass slightly higher than Jupiter's, $M_p = 1.19 \pm 0.12$ $M_{\rm Jup}$. 
The resulting orbital eccentricity is consistent with zero, $e = 0.05_{-0.03}^{+0.06}$. For WASP-72, assuming a stellar mass $M_{\ast} = 
1.23 \pm 0.10$ $M_{\sun}$  (Sect.~3.1), the fitted semi-amplitude $K = 179 \pm 6$ m\,s$^{-1}$ translates into a secondary mass  
$M_p = 1.31 \pm 0.08$ $M_{\rm Jup}$, while the deduced orbital eccentricity is also consistent with zero, $e = 0.05_{-0.03}^{+0.03}$. 

A model with a slope is slightly favored in the case of WASP-72, its value being -82 $\pm$ 22 \ms~per year. Indeed,
 the respective values for the Bayesian Information Criterion (BIC; Schwarz 1978)  led to likelihood ratios 
 (Bayes factors) between 10 and 55 in favor of 
 the slope model, depending if the orbit was assumed to be circular or not. Such  values for the
 Bayes factor are not high enough to be decisive, 
 and more RVs will be needed to confirm this possible trend.

To confirm that  the RV signal originates well from planet-mass objects orbiting the stars, 
we analyzed the {\tt CORALIE} cross-correlation functions (CCF) using the line-bisector technique described in Queloz et al (2001). 
The bisector spans revealed to be stable, their standard deviation being close to their average error (57 $vs$ 47 \ms for WASP-64 and 
28 $vs$ 24 \ms for WASP-72). No evidence for a correlation between the RVs and the bisector spans was found (Fig.~7 \& 8, lower panels), 
 the slopes deduced from linear regression being $-0.02 \pm 0.08$ (WASP-64) and $-0.01 \pm 0.04$ (WASP-72). These values and errors  
 makes any blend scenario very unlikely. Indeed, if the orbital signal of a putative blended eclipsing binary  (EB) is able to create a clear periodic 
 wobble of the sum of both CCFs,  it should also create a significant periodic distortion of its shape, resulting  in correlated variations of 
 RVs and bisector spans having the same order of magnitude (Torres et al. 2004). The power of this effect to identify blended EBs among transit
 candidates was first demonstrated by the classical case of  CSC 01944-02289 (Mandushev et al. 2005), for which the bisector spans
 varied in phase with the RVs and with an amplitude about twice lower.  Another famous case
 is the HD\,41004 system (Santos et al. 2002), with a K-dwarf blended with a M-dwarf companion 
(separation $\sim$0.5'') which  is itself orbited by a short-period brown dwarf. For this extreme system, the RVs show a clear signal at the period
of the brown dwarf orbit (1.3 days) and with an amplitude $\sim 50$ \ms that could be taken for the signal of a sub-Saturn mass planet 
orbiting the K-dwarf, except that the slope of the bisector-RV relation is $0.67 \pm 0.03$, clearly revealing that the main spectral component 
of the CCF is  not responsible for the observed signal. In the case of WASP-64 and 72, the 3-$\sigma$ upper limits of 0.23 and 0.09 that we derived
from Monte-Carlo simulations for the bisector-RV slopes combined with the much higher amplitude of the measured RV signals allow 
us to confidently infer that  the RV signal is actually originating from the target stars.  This conclusion is strengthened by the consistency
of the solutions derived from the global analysis of our spectroscopic and photometric data (see next Section).
We conclude thus that the stars WASP-64 and 
WASP-72 are transited by a giant planet  every $\sim$1.573 and $\sim$2.217 days, respectively. Of course,
 we cannot exclude that the light of those stars is not diluted by a well-aligned object able to bias our inferences about the planets. 
 Still, our multicolor transit photometry showing no dependance of the transit depths on the wavelength, and  the absence of any 
 detectable second spectra in the {\tt CORALIE}  data strongly disfavors any significant pollution of the light of the host stars.
 
\begin{figure}
\label{fig:rv}
\centering                     
\includegraphics[width=8.5cm]{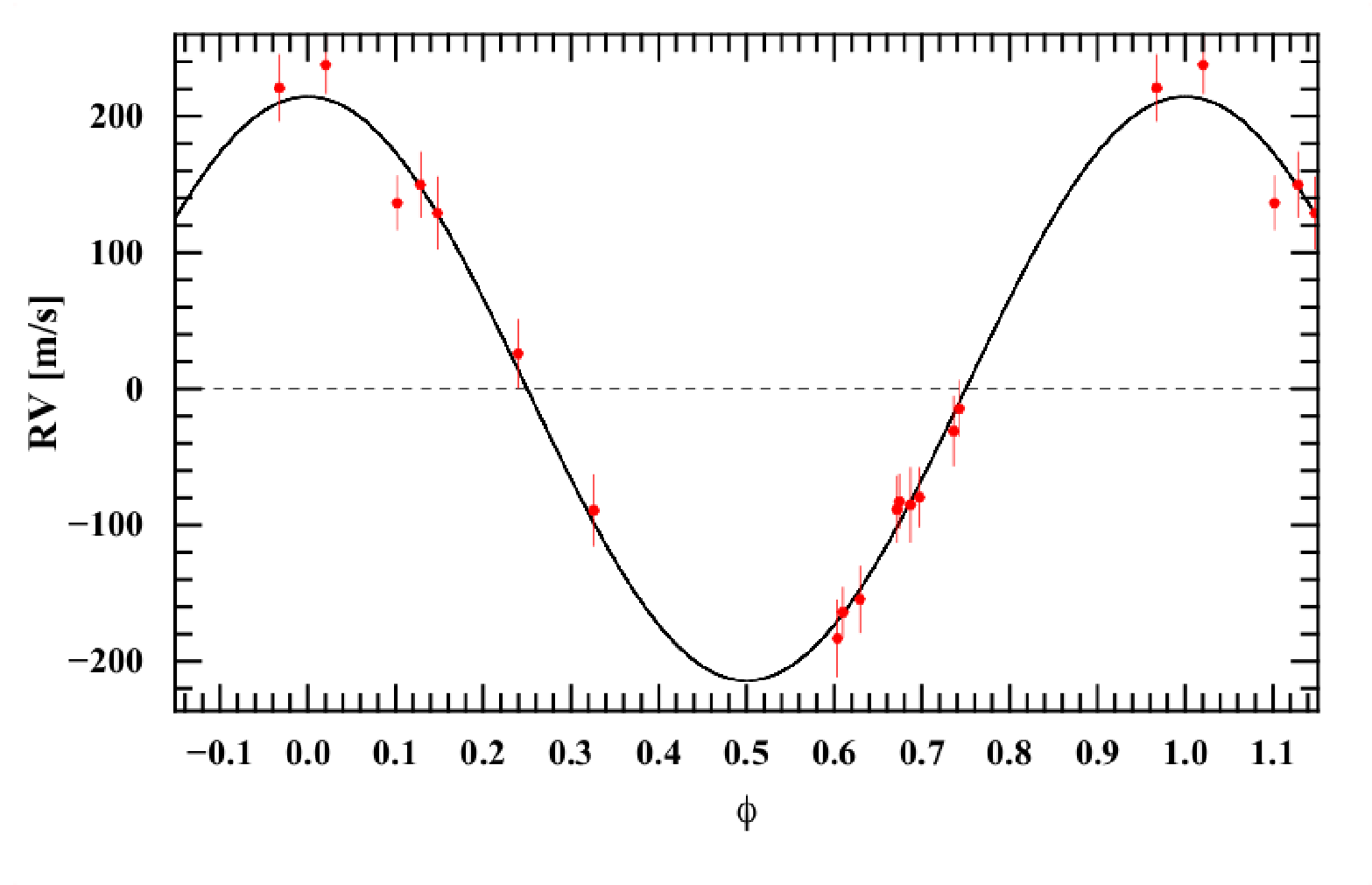}
\includegraphics[width=8.5cm]{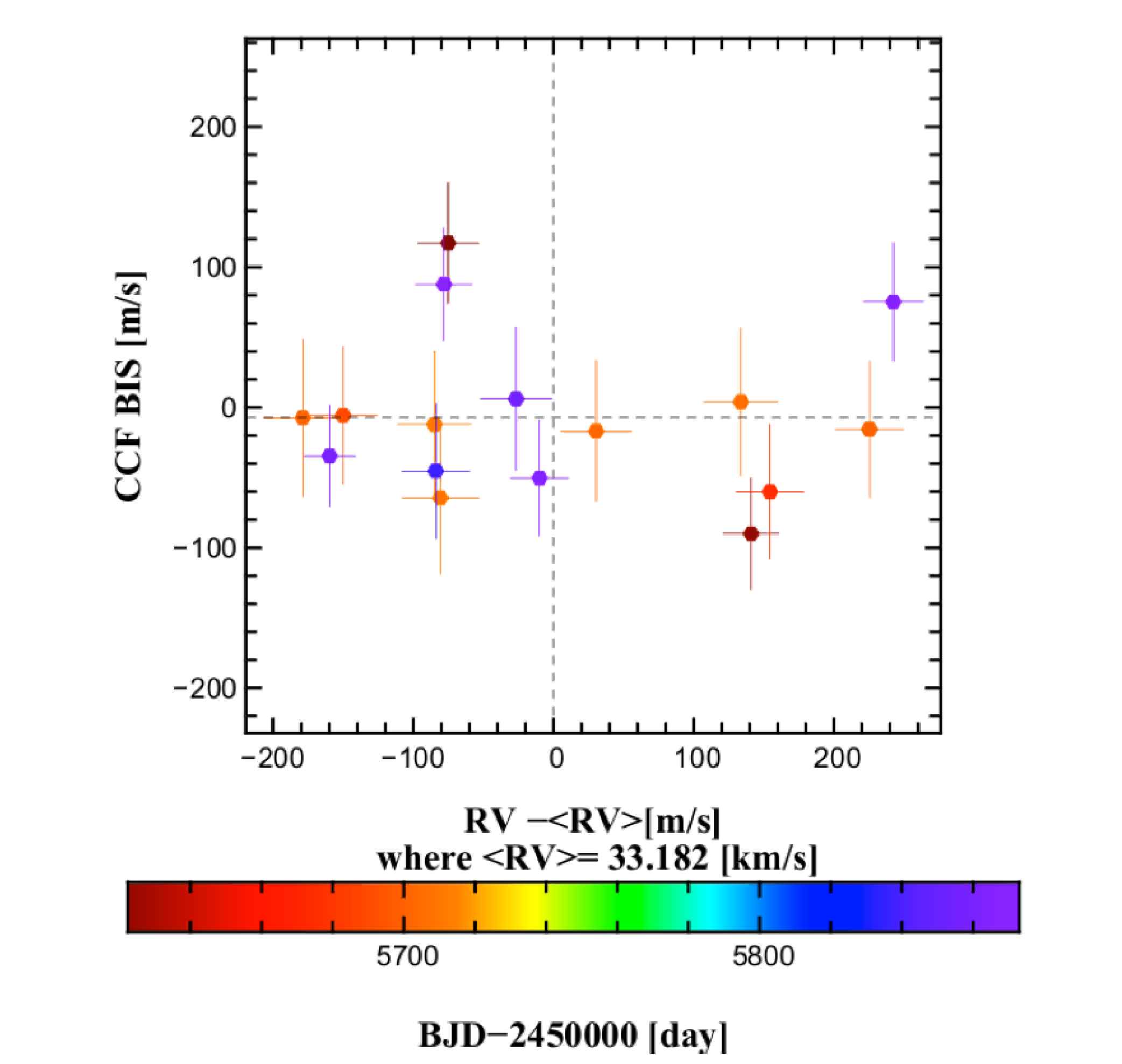}
\caption{$Top$: {\tt CORALIE} RVs for WASP-64 phase-folded on the best-fit orbital period, and with the best-fit
Keplerian model over-imposed. $Bottom$: correlation diagram CCF bisector spans $vs$ RV. The colors indicate the
measurement timings. }
\end{figure}

\begin{figure}
\label{fig:rv}
\centering                     
\includegraphics[width=8.5cm]{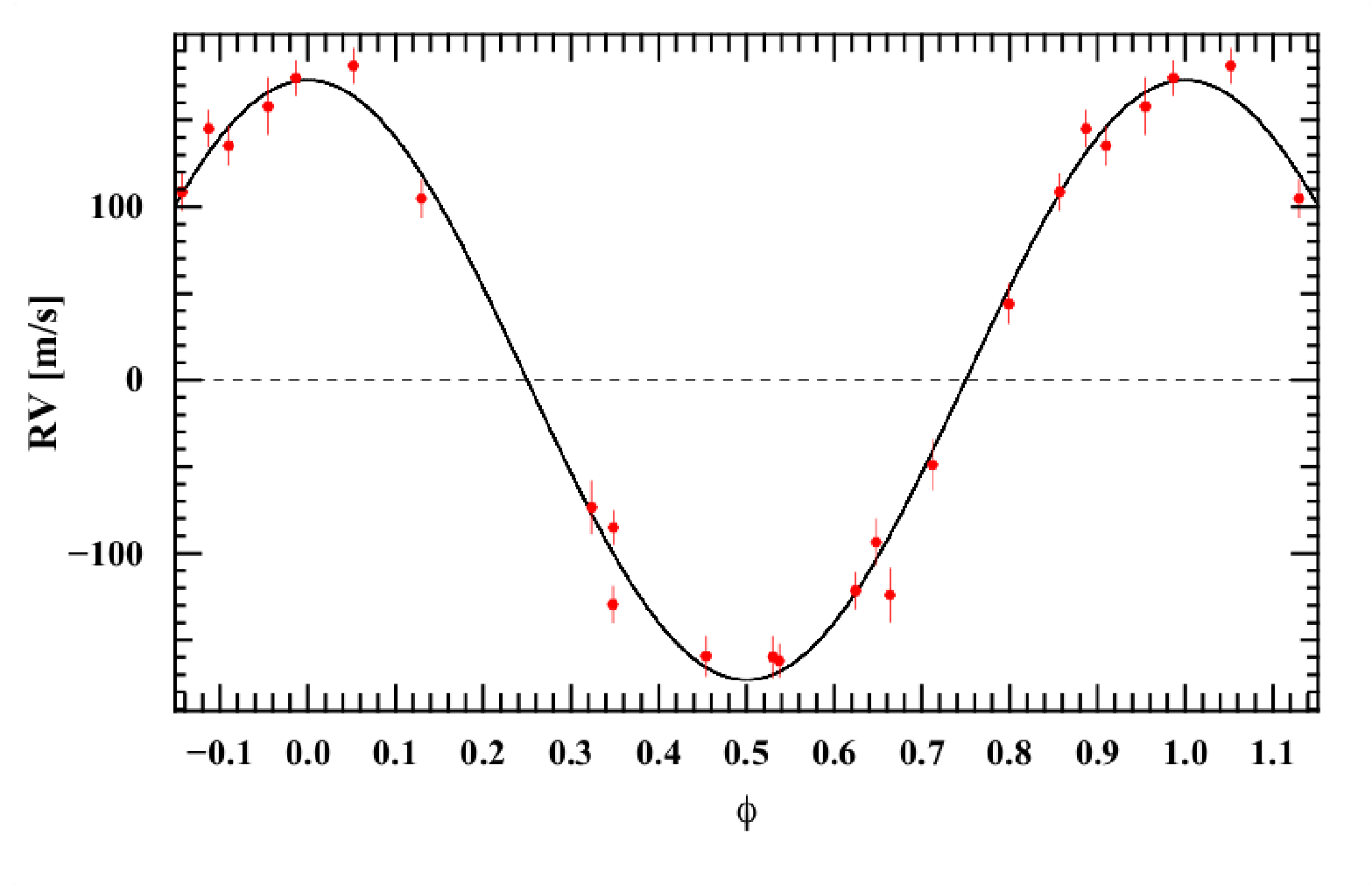}
\includegraphics[width=8.5cm]{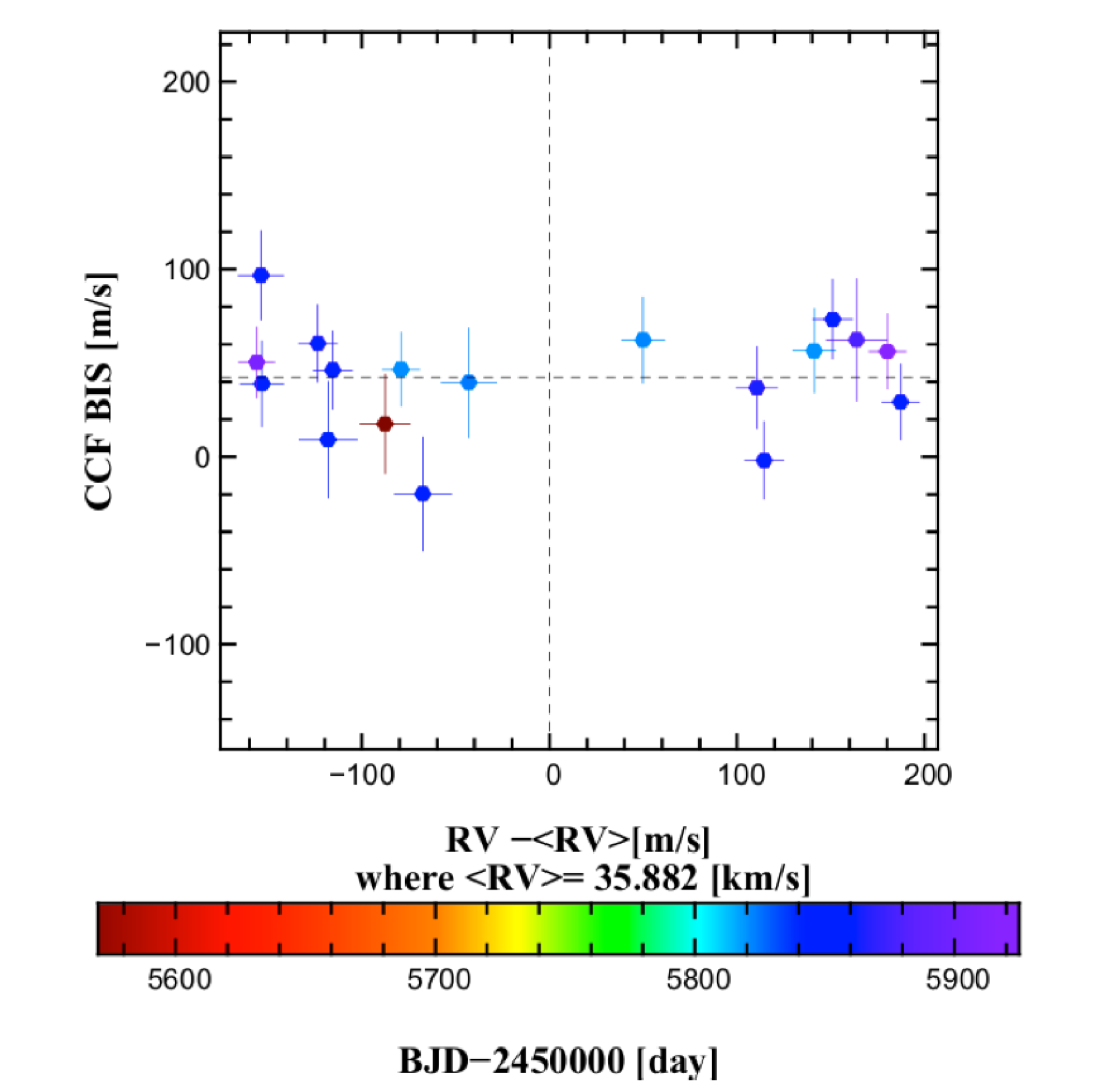}
\caption{$Top$: {\tt CORALIE} RVs for WASP-72 phase-folded on the best-fit orbital period, and with the best-fit
Keplerian model over-imposed. $Bottom$: correlation diagram CCF bisector spans $vs$ RV. The colors indicate the
measurement timings. }
\end{figure}

\begin{table}
\begin{center}
\begin{tabular}{ccccc}
\hline
Target & $HJD_{TDB}$-2450000 & RV & $\sigma_{RV}$ & BS\\ 
 &  & (km~s$^{-1}$) & (m~s$^{-1}$) & (km~s$^{-1}$) \\ \hline \noalign {\smallskip} 
WASP-64 & 5622.652685 & 33.1075 &  21.5 & 0.1171  \\ \noalign {\smallskip} 
WASP-64 & 5629.582507 & 33.3234 & 19.8 & --0.0900  \\ \noalign {\smallskip} 
WASP-64 & 5681.541466 & 33.3368 & 23.9 & --0.0601  \\ \noalign {\smallskip} 
WASP-64 & 5696.488635 & 33.0326 & 24.3 & --0.0057   \\ \noalign {\smallskip} 
WASP-64 & 5706.460152 & 33.4079 & 24.3 & --0.0156    \\ \noalign {\smallskip} 
WASP-64 & 5707.460947 & 33.0038 & 28.0 & --0.0075  \\ \noalign {\smallskip} 
WASP-64 & 5708.461813 & 33.2129 &  25.1 & --0.0169   \\ \noalign {\smallskip} 
WASP-64 & 5711.463695 & 33.3160 & 26.3 & 0.0040  \\ \noalign {\smallskip} 
WASP-64 & 5715.458133 & 33.1019 & 27.2 & --0.0645  \\ \noalign {\smallskip} 
WASP-64 & 5716.463634 & 33.0977 & 26.1 & --0.0121  \\ \noalign {\smallskip} 
WASP-64 & 5842.867334 & 33.0986 & 24.0 &  --0.0453  \\ \noalign {\smallskip} 
WASP-64 & 5861.847564 & 33.1559 & 25.4  & 0.0061  \\ \noalign {\smallskip} 
WASP-64 &  5864.795499 & 33.0230 & 18.1 &  --0.0346   \\ \noalign {\smallskip} 
WASP-64 &  5869.724123 & 33.1725 & 20.6 & --0.0504  \\ \noalign {\smallskip} 
WASP-64 & 5871.734734 & 33.4248 & 21.1 &  0.0752   \\ \noalign {\smallskip} 
WASP-64 & 5872.762875 & 33.1044 & 20.1 &  0.0877   \\[0.33cm]
WASP-72 & 5570.618144 & 35.7945 & 13.3 &	0.0176    \\ \noalign {\smallskip} 
WASP-72 & 5828.890931	 & 36.0233 & 11.3 & 0.0566     \\ \noalign {\smallskip} 
WASP-72 & 5829.865219	 & 35.8029 & 9.8  &  0.0466      \\ \noalign {\smallskip} 
WASP-72 & 5830.866138	 & 35.9320 & 11.5 & 0.0623      \\ \noalign {\smallskip} 
WASP-72 & 5832.894502	 & 35.8391 & 14.6 & 0.0396	 \\ \noalign {\smallskip} 
WASP-72 & 5852.780149	 & 35.7640 & 15.5 & 0.0092	\\ \noalign {\smallskip} 
WASP-72 & 5856.757447	 & 35.7288 & 11.5 & 	0.0390	\\ \noalign {\smallskip} 
WASP-72 & 5858.689011	 & 35.8146 & 15.2 &	--0.0198	\\ \noalign {\smallskip} 
WASP-72 & 5863.800340	 & 35.7664 & 10.4 & 	0.0462	\\ \noalign {\smallskip} 
WASP-72 & 5864.750293	 & 36.0695 & 10.0 & 0.0292	\\ \noalign {\smallskip} 
WASP-72 & 5865.812141	 & 35.7283 & 11.9 & 0.0968	\\ \noalign {\smallskip} 
WASP-72 & 5866.605265	 & 36.0331 & 10.6 & 	0.0734	\\ \noalign {\smallskip} 
WASP-72 & 5867.628720	 & 35.7586 & 10.3 &	0.0605	\\ \noalign {\smallskip} 
WASP-72 & 5868.759336	 & 35.9966 & 10.4 & --0.0019	\\ \noalign {\smallskip} 
WASP-72 & 5873.808914	 & 35.9928 & 10.9 &	0.0368	\\ \noalign {\smallskip} 
WASP-72 & 5886.748911	 & 36.0460 & 16.3 &  0.0624 \\ \noalign {\smallskip} 
WASP-72 & 5914.700813	 & 35.7261  & 9.5 & 	0.0504	\\ \noalign {\smallskip} 
WASP-72 & 5924.584434	 & 36.0624 & 10.0 &	0.0562	\\ \noalign {\smallskip} 
\hline
\end{tabular}
\caption{{\tt CORALIE} radial-velocity measurements for WASP-64 and WASP-72 (BS = bisector spans).}
\end{center}
\label{wasp64+72-rvs}
\end{table}

\section{Analysis}

\subsection{Spectroscopic analysis - stellar properties}

The {\tt CORALIE} spectra of WASP-64 and WASP-72 were co-added to produce
single spectra with average S/N of 60 and 80, respectively. The standard pipeline
reduction products were used in the analysis.

The spectral analysis was performed using the methods given by
Gillon et al. (2009a). The \halpha\ line was used to determine the
effective temperature (\teff). For WASP-64,  the Na\,{\sc i}\,D and Mg\,{\sc i}\,b lines
were used as surface gravity ($\log g$) diagnostics.  For WASP-72,
getting an measurement of $\log g$ was more critical, as
the transit photometry does not constrain strongly the stellar density (see Sec. 3.2), so
we used the improved method recently described by Doyle et al. (2013) and based on the ionization balance
of selected Fe~{\sc i}/Fe~{\sc ii} lines in addition to the pressure-broadened Ca~{\sc i} lines at 6162{\AA} and 6439{\AA}
 (Bruntt et al. 2010a), along with the Na~{\sc i} D lines. 
 The parameters 
obtained from the analysis are listed in Table~3. 
The elemental abundances were determined from equivalent width measurements 
of several clean and unblended lines. A value for microturbulence (\mictrb) 
was determined from Fe~{\sc i} lines using the method of Magain (1984). The 
quoted error estimates include those given by the uncertainties in \teff, \logg\ 
and \mictrb, as well as the scatter due to measurement and atomic data uncertainties.

The projected stellar rotation velocities (\vsini) were determined by fitting the
profiles of several unblended Fe~{\sc i} lines. Values for macroturbulence
(\mactrb) of 1.8 $\pm$ 0.3 and 4.0 $\pm$ 0.3 {\kms} were assumed for 
WASP-64 and WASP-72, respectively, based on the calibration by
Bruntt at al. (2010b). An instrumental FWHM of 0.11 $\pm$ 0.01~{\AA} was
determined for both stars from the telluric lines around 6300\AA. 
Best-fitting values of \vsini\ = 3.4 $\pm$ 0.8 \kms\ (WASP-64) and 
\vsini\ = 6.0 $\pm$ 0.7 \kms\ (WASP-72) were obtained.

\begin{table}[h]
\begin{center}
\begin{tabular}{ccc} \hline  
Parameter  & WASP-64 & WASP-72 \\ \hline 
RA (J2000)  & 06 44 27.61 & 02 44 09.60 \\
DEC (J2000) & -32 51 30.25 & -30 10 08.5 \\
$V$ & 12.29 & 10.88 \\
$K$ &10.98  & 9.62 \\
\teff      &   5550 $\pm$ 150 K  & 6250 $\pm$ 100 K \\
\logg      &   4.4 $\pm$ 0.15  &  $4.08 \pm 0.13$ \\
\mictrb    &   0.9 $\pm$ 0.1 \kms  &   1.6 $\pm$ 0.1 \kms \\
\vsini     &   3.4 $\pm$ 0.8 \kms &   6.0 $\pm$ 0.7 \kms \\
{[Fe/H]}   &$-$0.08 $\pm$ 0.11 &$-$0.06 $\pm$ 0.09 \\
{[Na/H]}   &   0.14 $\pm$ 0.08 &$-$0.03 $\pm$ 0.04 \\ 
{[Mg/H]}   &   0.12 $\pm$ 0.12    &  \\
{[Al/H]}   &    0.00 $\pm$ 0.08 &     \\
{[Si/H]}   &    0.10 $\pm$ 0.10 &   0.02 $\pm$ 0.07 \\
{[Ca/H]}   &   0.05 $\pm$ 0.16 &  0.07 $\pm$ 0.14 \\
{[Sc/H]}   &    0.07 $\pm$ 0.10 & 0.14 $\pm$ 0.07 \\
{[Ti/H]}   &   $-$0.02 $\pm$ 0.11 & 0.07 $\pm$ 0.11 \\
{[V/H]}    &   0.03 $\pm$ 0.16 &$-$0.01 $\pm$ 0.08 \\
{[Cr/H]}   &   0.01 $\pm$ 0.08 &   0.02 $\pm$ 0.10 \\
{[Mn/H]}   &   0.09 $\pm$ 0.10 &$-$0.11 $\pm$ 0.06 \\
{[Co/H]}   &   0.06 $\pm$ 0.09 &$-$0.06 $\pm$ 0.18 \\
{[Ni/H]}   &$-$0.04 $\pm$ 0.11 &$-$0.04 $\pm$ 0.06 \\
log A(Li)  &   $<$ 0.61 $\pm$ 0.15  & $<$1.21 $\pm$ 0.17 \\
Mass       &   0.98 $\pm$ 0.09 $M_{\sun}$ &   $1.31 \pm 0.11 M_{\sun}$ \\
Radius     &   1.03 $\pm$ 0.20 $R_{\sun}$ &  $1.72 \pm 0.31 R_{\sun}$ \\
Sp. Type   &   G7 & F7 \\
Distance   &   350 $\pm$ 90 pc  &   $340 \pm 60$ pc \\ \hline 
\end{tabular}
\end{center}
\label{wasp64+72-params}
\caption{Basic and spectroscopic parameters of WASP-64 and WASP-72
from spectroscopic analysis. \newline
{\bf Notes:} The values for the stellar mass, radius and 
surface gravity are given here for information purpose only. The values that
we finally adopted for these parameters are the ones derived
from the global analysis of our data (Sec. 3.2) and are presented in Table 5 and 6.
Mass and radius estimate using the
calibration of Torres et al. (2010). Spectral type estimated from \teff\
using the table in Gray (2008). } 
\end{table}

There is no significant detection of lithium in the spectra, with equivalent
width upper limits of 2m\AA  ~for both stars, corresponding to abundance 
upper limits of log~A(Li) $<$ 0.61 $\pm$ 0.15 (WASP-64) and log~A(Li) 
$<$ 1.21 $\pm$ 0.17 (WASP-72). These imply ages of at least a few Gyr 
(Sestito \& Randich, 2005).

The rotation rate for WASP-64 ($P_{\rm rot} = 15.3 \pm 4.7$~d) 
and WASP-72 ($P_{\rm rot} = 14.5 \pm 3.1$~d) implied by the {\vsini} 
give gyrochronological ages of $\sim$$1.2_{-0.7}^{+1.2}$ Gyr (WASP-64) and
$\sim 3.7_{-1.9}^{+4.0}$ Gyr (WASP-72) under the Barnes (2007) relation.
 
We obtained with {\tt CORALIE} two spectra of TYC7091-1288-1, the brighter star 
lying at 28'' East from WASP-64 (Sec. 2.2.1, Fig.~3).  The co-added spectrum has a S/N
 of only $\sim$30. A spectral analysis led to \teff $\sim$5700K and \logg $\sim$4.5, 
 with no sign of any significant lithium absorption, and a low {\vsini} $\sim$4~\kms. The RV is $\sim$35 {\kms}, 
compared to 33.2~{\kms} for WASP-64. The cross-correlation function reveals that 
TYC 7091-1288-1 is an SB2 system (Fig.~8). The PPMXL catalogue
(Roeser et al. 2010) shows that proper motions of both stars are consistent to within their quoted
uncertainties. If these stars are physically associate, as suggested by their similar proper motions
and  radial velocities, their angular separation corresponds to a projected distance of 9800 $\pm$ 2500 AU, 
which is possible for a very wide triple system. Of course, the spectra of TYC7091-1288-1 and 
WASP-64 are totally separated at the spatial resolution of  {\tt CORALIE} (typical seeing $\sim$1'', 
fiber diameter of 2''), considering the 28'' separation between both objects.
 
\begin{figure}
\label{fig:rv}
\centering                     
\includegraphics[width=9cm,angle=0]{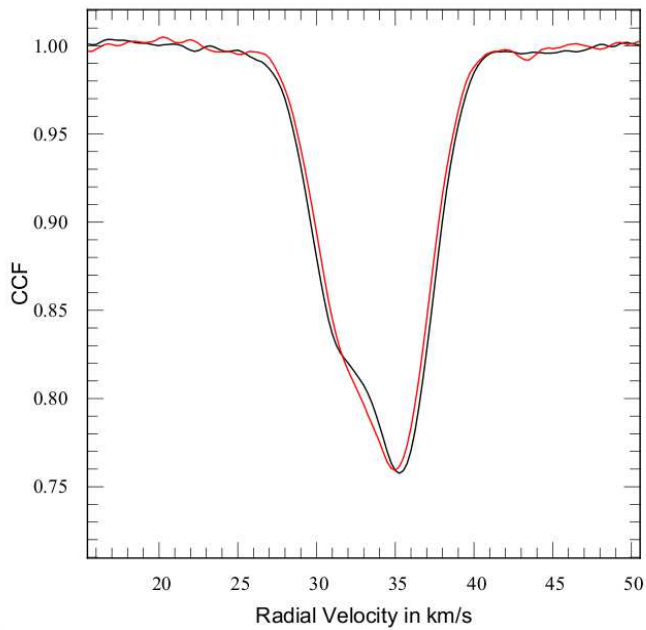}
\caption{The cross-correlation functions  for the two {\tt CORALIE} spectra of TYC7091-1288-1.
Their clear asymmetry indicates  the SB2 nature of the star.}
\end{figure}
 
\subsection{Global analysis}

For both systems, we performed a global analysis of the follow-up photometry and the
 {\tt CORALIE} RV measurements.  The analysis was performed using the adaptive 
Markov Chain Monte-Carlo (MCMC) algorithm described by Gillon et al. (2012, and references therein). 
To summarize, we simultaneously fitted the data, using for the photometry the transit 
model of Mandel \& Agol (2002) multiplied  by a baseline model consisting of a different 
second-order polynomial in time for each of the light curves. We motivate the choice of this 
`minimal' baseline model by its better ability to represent properly any smooth variation due to a combination of
differential extinction and low-frequency stellar variability, compared to other possible simple functions
 (scalar, linear function of time or airmass). We outline that using a simple scalar as baseline model
relies on the strong assumptions that the ensemble of comparison stars used to derive the differential 
photometry has exactly the same color than the target, that the target is perfectly stable, and that no low-frequency
noise could have modified the transit shape.
For  eight TRAPPIST light curves 
(see Table 1), a normalization offset was also part of the model to represent the effect of
 the meridian flip. TRAPPIST's mount is a german equatorial type, which means
  that the telescope has to undergo a 180$^{\circ}$ rotation when the meridian is reached, 
  resulting in different locations of the stellar images on the detector before and after the flip,
  and thus in a possible jump of the differential flux at the time of the flip.  For the first WASP-72 {\it Euler} 
  transit, a quadratic function of airmass had to be added to the minimal baseline model to account
  for the strong extinction effect caused by the high airmass ($>$2.5) at the end of the run.
For the second WASP-72 transit observed by {\it Euler}, a normalization offset and a linear term in $x$- and 
$y$-position were added to model the effects on the photometry of the telescope tracking problem.
 On their side, the RVs were  represented by a classical model assuming Keplerian orbits 
 (e.g. Murray \& Correia 2010, eq. 65), plus a linear trend for WASP-72  (see Sec.~2.3).

The jump parameters\footnote{Jump parameters are the parameters that are randomly 
perturbed at each step of the MCMC.} in our MCMC analysis were: the planet/star area 
ratio $(R_p /R_\ast )^2$, the transit width (from first to last contact) $W$,  the parameter $b' = 
a \cos{i_p}/R_\ast$ (which is the transit impact parameter in case of a circular orbit) where
$a$ is the planet's semi-major axis and $i_p$ its orbital inclination,  the orbital 
period $P$ and time of minimum light $T_0$,  the parameter $K_2 = K  \sqrt{1-e^2}  
 \textrm{ } P^{1/3}$, where $K$ is the RV orbital semi-amplitude,  and the two parameters $\sqrt{e} \cos{\omega}$ 
and $\sqrt{e} \sin{\omega}$, where $e$ is the orbital eccentricity and $\omega$ is the argument 
of periastron. The choice of these two latter parameters is motivated by our will to avoid biasing
the derived posterior distribution of $e$,  as their use corresponds to a uniform prior for $e$ (Anderson
et al. 2011).

We assumed a uniform prior distribution for all these jump parameters. 
The photometric baseline model parameters and the systemic radial velocity of the star $\gamma$ (and the
 slope of the trend for WASP-72) were not actual jump parameters; they were determined by 
 least-square minimization at each step of the MCMC. 

We assumed a quadratic limb-darkening law, and we allowed  the quadratic coefficients $u_1$ 
and $u_2$ to float in our MCMC analysis, using as jump parameters not these coefficients 
themselves but the combinations $c_1 = 2 \times u_1 + u_2$  and $c_2 = u_1 - 2 \times u_2$ 
to minimize the correlation of the uncertainties (Holman et al. 2006).  To obtain a 
limb-darkening solution consistent with theory, we used normal prior distributions 
for $u_1$ and $u_2$ based on theoretical values and 1-$\sigma$ errors interpolated in the 
tables by Claret (2000; 2004) and shown in Table 4. For our two non-standard filters ($I+z$ and 
blue-blocking), we estimated the effective wavelength basing on the transmission curves
of the filters, the quantum efficiency curve of the camera and the spectral energy distributions 
of the stars (assumed to emit as blackbodies), and we interpolated the corresponding 
limb-darkening coefficient values in Claret's tables and estimated their errors by using the values for the
two nearest standard filters.  We tested the insensitivity of our results to the details of this 
interpolation by performing short MCMC analyses with different prior distributions for the limb-darkening
coefficients of the non-standard filters (e.g. assuming $I+z$ = $I$-Cousins, blue-blocking = $R$-Cousins, etc.) 
which led to results fully consistent with those of our nominal analysis. Such tests had also been performed in
the past for other WASP planets, with similar results (e.g. Smith et al. 2012).

\begin{table}
\begin{center}
\begin{tabular}{ccc}
\hline \noalign {\smallskip}
{\it Limb-darkening coefficient} &  WASP-64 & WASP-72 \\ \noalign {\smallskip}
\hline \noalign {\smallskip}
$u1_{V}$            &  $0.50 \pm 0.035$   &    -                                \\ \noalign {\smallskip} 
$u2_{V}$            &  $0.23 \pm 0.025$   &     -                                 \\ \noalign {\smallskip} 
$u1_{Gunn-r}$   &  $0.43 \pm 0.03$ &   $0.31 \pm 0.015$    \\ \noalign {\smallskip} 
$u2_{Gunn-r}$     &  $0.26 \pm 0.02$ &   $0.315 \pm 0.005$  \\ \noalign {\smallskip} 
$u1_{BB}$          &  $0.36 \pm 0.05$   &   $0.255 \pm 0.04$   \\ \noalign {\smallskip} 
$u2_{BB}$          &  $0.255 \pm 0.02$   &   $0.305 \pm 0.01$   \\ \noalign {\smallskip} 
$u1_{I+z}$          &  $0.29 \pm 0.03$   &   $0.205 \pm 0.02$   \\ \noalign {\smallskip} 
$u2_{I+z}$          &  $0.255 \pm 0.015$ &   $0.295 \pm 0.01$  \\ \noalign {\smallskip} 
\hline \noalign {\smallskip}
\end{tabular}
\caption{Expectation and standard deviation of the normal distributions used
as prior distributions for the quadratic limb-darkening coefficients $u1$ and $u2$ in our 
MCMC analysis. }
\end{center}
\end{table}

Our analysis was composed of five Markov chains of $10^5$ steps, the first 20\% of each 
chain being considered as its burn-in phase and discarded. For each run the convergence of the 
five Markov chains was checked using the statistical test presented by Gelman and Rubin (1992). 
The correlated noise present in the light curves was taken into account as described by  Gillon et al. (2009b),
by comparing the scatters of the residuals in the original and in 
time-binned versions of the data, and by rescaling the errors accordingly. 
For the WASP-72 RVs, a  `jitter' noise of 5.1 \ms\  was added quadratically to the error bars, 
to equalize the mean error with the $rms$ of the best-fitting model residuals.

At each step of the Markov chains the dynamical stellar density $\rho_\ast$ deduced from the 
jump parameters (b', W, $(R_p /R_\ast )^2$, $\sqrt{e} \cos{\omega}$, $\sqrt{e} \sin{\omega}$, $P$; 
see, e.g., Winn 2010)
and values for  $T_{\rm eff}$ and  [Fe/H] drawn from the normal distributions deduced from our 
spectroscopic analysis (Sect.~3.1), were used to determine a value for the stellar mass $M_\ast$
 through an empirical law $M_\ast$($\rho_\ast$, $T_{\rm eff}$, [Fe/H]) (Enoch et al. 2010; 
 Gillon et al. 2011) calibrated using the parameters of the extensive list of stars belonging to
 detached eclipsing binary systems presented by Southworth (2011). For WASP-64, the list
 was restricted to the 113 stars with a mass between 0.5 and 1.5 $M_\odot$, while the
 212 stars with a mass between 0.7 to 1.7 $M_\odot$ were used for WASP-72, the goal 
 of this selection being to benefit from our preliminary estimation of the stellar mass  (Sec. 3.1, Table 3)
 to improve the determination of the physical parameters while using a number 
 of calibration stars large enough to avoid small number statistical effects. To propagate properly the 
 errors on the calibration law, the parameters of the selected subset of eclipsing binary stars 
 were normally perturbed within their observational error bars and the coefficients of the law were 
 redetermined at each MCMC step. Using the resulting stellar mass, the physical parameters 
 of the system were then deduced from the jump parameters.  In this procedure to derive
 the physical parameters of the system, the  spectroscopic stellar gravity is thus not used, 
 the stellar density deduced from the dynamical + transit parameters constraining by itself 
 the evolutionary state of the star (Sozzetti et al. 2007). Still, for WASP-72 we assumed a normal
 prior distribution for  \logg  based on the spectroscopic value and error bar (Table 3), because 
 the low transit depth combined with the significant level of correlated noise of our data led to 
 relatively poor constraint on the stellar density from the photometry alone (error of 50\%).
 
 For both systems, two analyses were performed, one assuming a circular orbit and the other
 an eccentric orbit. For the sake of completeness, the derived parameters for both models are shown 
 in Table 5 (WASP-64) and Table 6 (WASP-72), while the best-fit transit models
 are shown in Fig.~10 and 11 for the circular model. Using the BIC as proxy for the
 model marginal likelihood, the resulting Bayes factors are $\sim$3000 (WASP-64)
 and $\sim$5000 (WASP-72) in favor of the circular models.  A circular orbit is
 thus favored for both systems, and we adopt the corresponding results
 as our nominal solutions (right columns of Table 5 and 6). This choice is strengthened by the modeling
 of the tidal evolution of both planets, as discussed in Sec. 4.
       
\subsubsection{Stellar evolution modeling}

After the completion of the MCMC analyses described above, we performed for both systems a stellar evolution 
modeling based on the code CLES (Scuflaire et al. 2008) and on the Levenberg-Marquardt optimization algorithm (Press et al. 1992), 
using as input the stellar densities deduced from the
 MCMC, and the effective temperatures and metallicities deduced from our 
spectroscopic analyses, with the aim  to assess the reliability of the deduced physical parameters and to estimate
the age of the systems. The resulting  stellar masses were  $0.95 \pm 0.05$ $M_\odot$
(WASP-64) and $1.34 \pm 0.11$ $M_\odot$ (WASP-72), consistent with the 
MCMC results, while the resulting ages were $7.0 \pm 3.5$ (WASP-64) and $3.2 \pm 0.6$ 
Gy (WASP-72). 

Unlike WASP-64, WASP-72 appears to be significantly evolved. To check further
the reliability of our inferences for the system, we derived its parameters using the solar calibrated value of 
the mixing length parameter and a value 20\% lower, and we also investigated the effects of convective
 core overshooting and microscopic diffusion of helium.  All the results are within 1sigma 
 for the mean density and surface metallicity, and within 1.5 sigma for the effective temperature, 
 however the best fits of mean density and $T_{eff}$  are found for models including convective 
 core overshooting.  Solutions with standard physics tend to produce mean density higher than 
 0.2 and $T_{eff}$ higher than 6300K.

\subsubsection{Global analysis of the transits with free timings}

 As a complement to our global analysis, we performed for both systems
 another global analysis with the timing of each transit being free parameters in the MCMC. 
The goal here was to benefit from the strong constraint brought on the transit shape provided by the total 
data set to derive accurate transit timings  and to assess the transit periodicity.  In this analysis, 
the parameters $T_0$ and $P$ were kept under the control of  normal prior distributions based on
the values shown for a circular orbit in Table 5 (WASP-64) and Table 6 (WASP-72), and we added
a timing offset as jump parameter for each transit. The orbits were assumed to be circular. The 
resulting transit timings and their errors are shown in Table 7. This table also shows (last column) the resulting
 transit timing variations (TTV = observed minus computed timing, O-C). These TTVs are shown as a function of the transit  epochs in Fig.~12.
 They are all compatible with zero, i.e. there is no sign of transit aperiodicity. 

\begin{figure}
\label{fig:cw64}
\centering                     
\includegraphics[width=8.5cm]{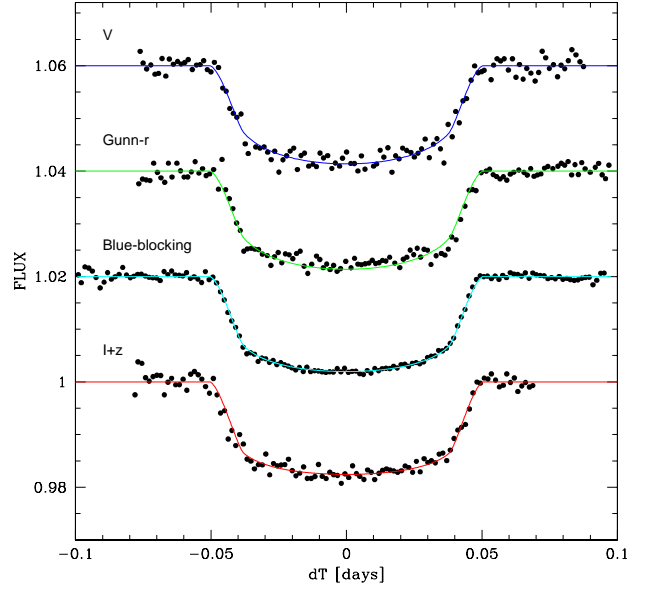}
\caption{Combined follow-up transit photometry for WASP-64\,b, detrended, period-folded and binned per
intervals of 2 min.  For each
filter, the best-fit transit model from the global MCMC analysis is superimposed. The $V$, Gunn-$r$ and 
blue-blocking light curves are shifted along the $y$-axis for the sake of clarity.}
\end{figure}

\begin{figure}
\label{fig:cw72}
\centering                     
\includegraphics[width=8.5cm]{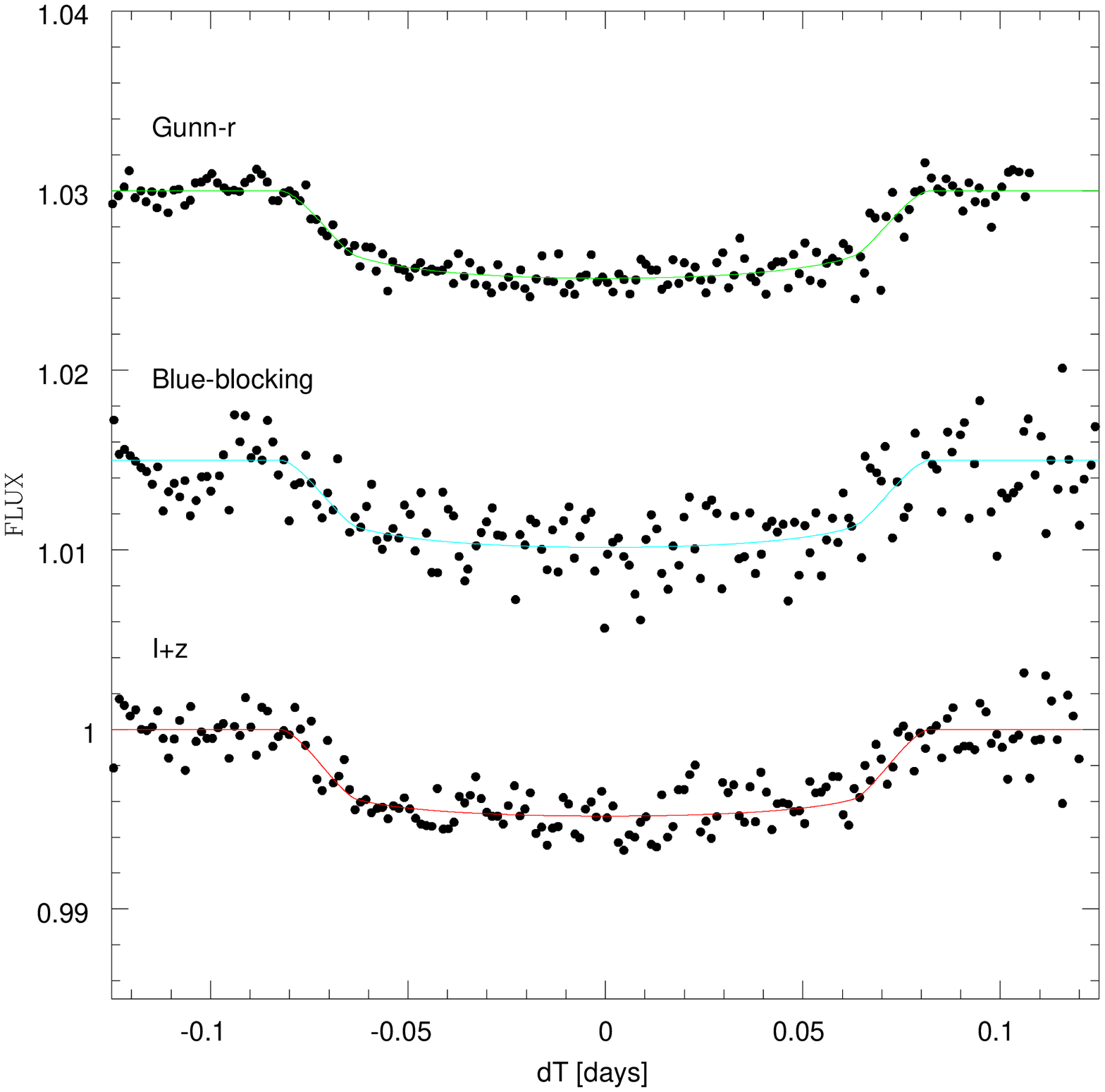}
\caption{Combined follow-up transit photometry for WASP-72\,b, detrended, period-folded and binned per
intervals of 2 min.  For each
filter, the best-fit transit model from the global MCMC analysis is superimposed. The Gunn-$r$ and 
blue-blocking light curves are shifted along the $y$-axis for the sake of clarity.}
\end{figure}

\begin{figure}
\label{fig:ttv}
\centering                     
\includegraphics[width=8.5cm]{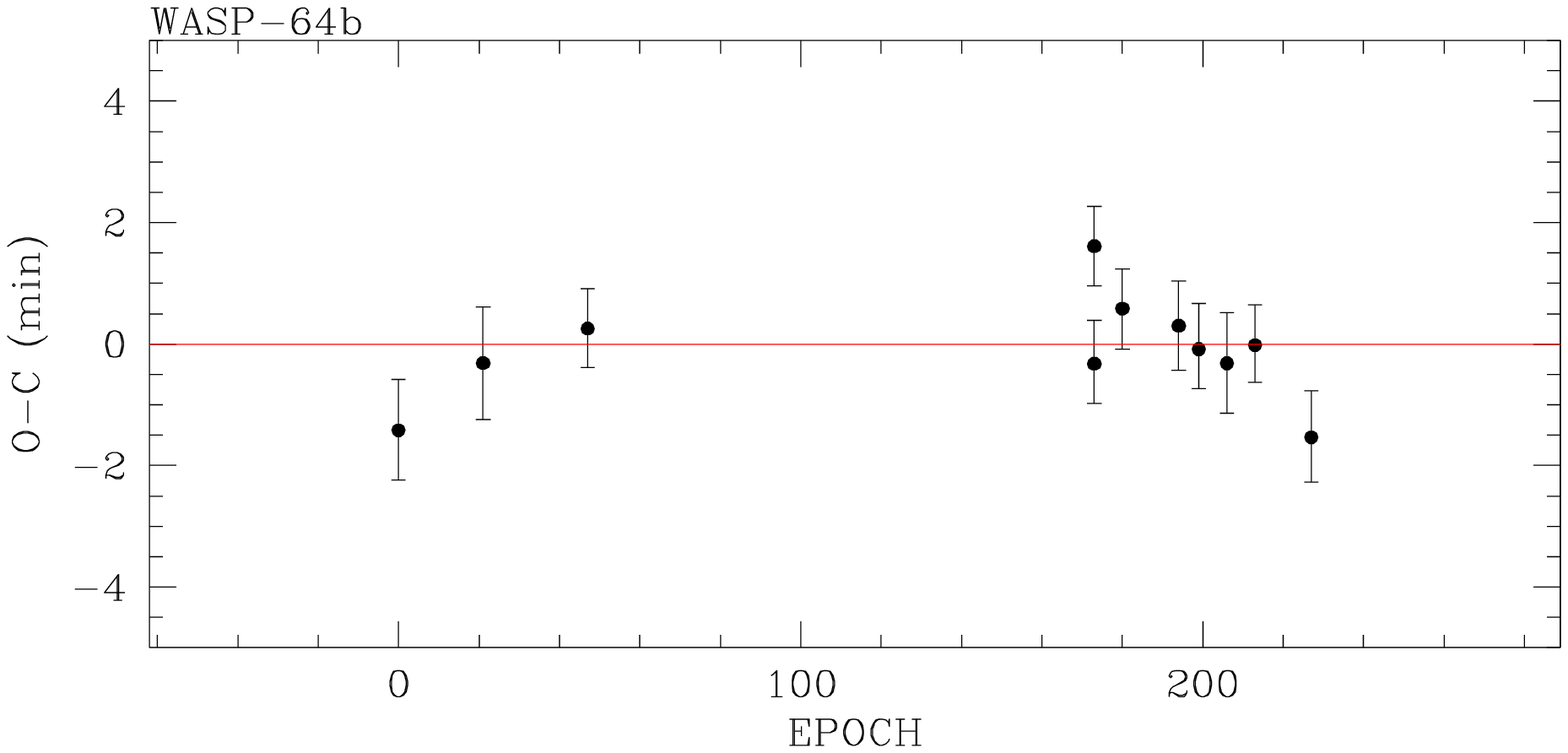}
\includegraphics[width=8.5cm]{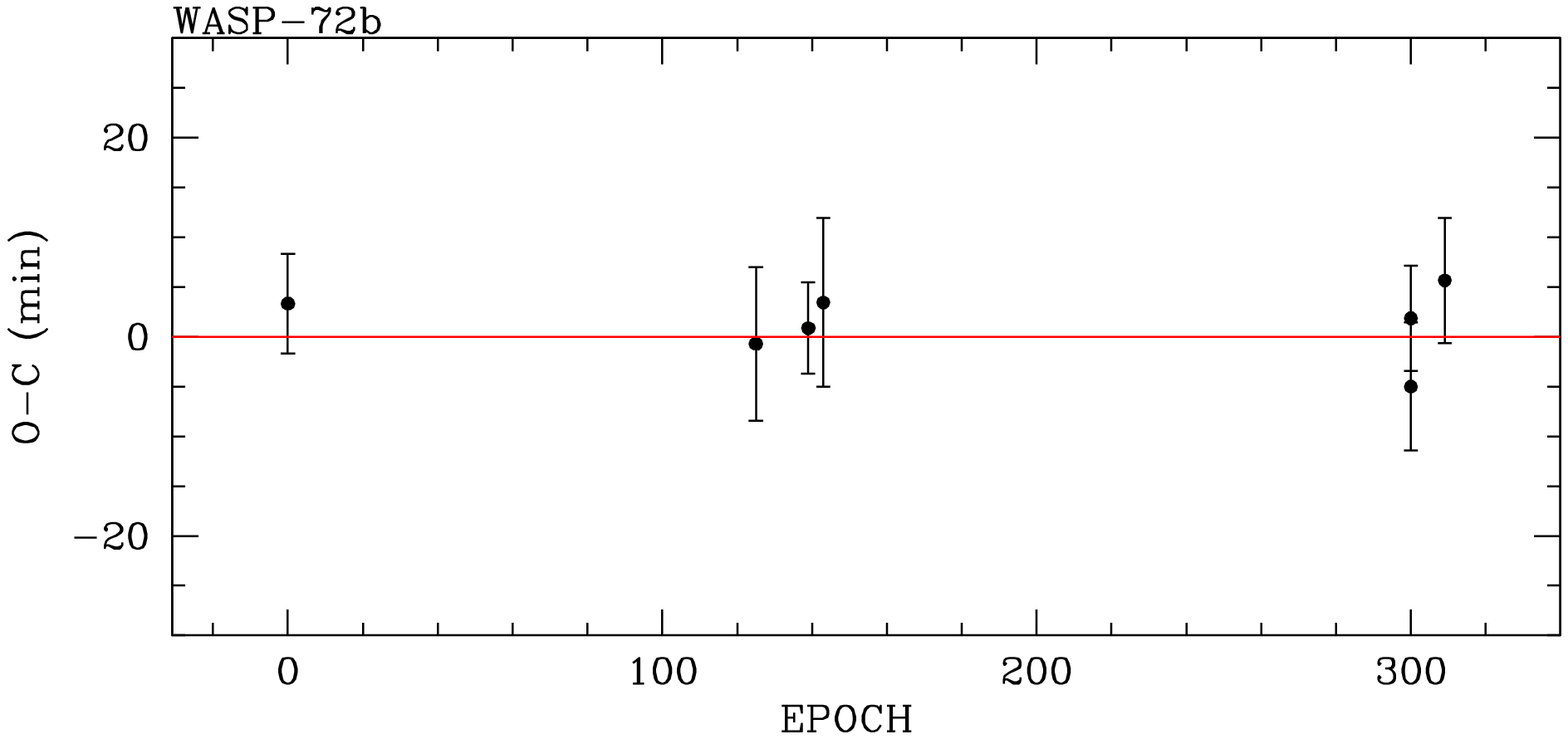}
\caption{TTVs ($O-C$ = observed minus computed timing) derived from the global MCMC analyzes of 
the WASP-64\,b (top) and WASP-72\,b (bottom) transits (see Sec. 3.2.1).}
\end{figure}

\begin{table*}[h]
\begin{center}
\begin{tabular}{ccc}
& {\bf WASP-64}  & \\
\hline \noalign {\smallskip}
$Free$ $parameters$ &  $e \geq 0 $ & $e = 0$ {\bf (adopted)} \\ \noalign {\smallskip}
\hline \noalign {\smallskip}
Planet/star area ratio  $ (R_p/R_\ast)^2 $  [\%]         &  $1.524 \pm 0.025$                       & $1.522 \pm 0.025$ \\ \noalign {\smallskip} 
$b'=a\cos{i_p}/R_\ast$ [$R_*$]                                    &  $0.321_{-0.065}^{+0.046}$        &   $0.322_{-0.068}^{+0.048}$                    \\ \noalign {\smallskip} 
Transit width  $W$ [d]                                                     &   $0.10005 \pm 0.00047$            & $0.09999 \pm 0.00046$                 \\ \noalign {\smallskip} 
$T_0-2450000$ [$HJD_{TDB}$]                                  &  $5582.60171 \pm 0.00026$      &$5582.60169 \pm 0.00027$ \\ \noalign {\smallskip}
Orbital period  $ P$ [d]                                                    &  $1.5732918 \pm 0.0000015$   &  $1.5732918 \pm 0.0000015$                 \\ \noalign {\smallskip} 
RV $K_2$  [m\,s$^{-1}$\,d$^{1/3}$]                              & $248 \pm 14$                                &    $257 \pm 11$                     \\ \noalign {\smallskip} 
RV $\gamma$ [km\,s$^{-1}$]                                         & $33.191 \pm 0.010$                    &  $33.1854 \pm 0.0057$   \\ \noalign {\smallskip} 
$\sqrt{e}\cos{\omega}$                                                   &  $0.083_{-0.088}^{+0.081}$       &    0 (fixed)             \\ \noalign {\smallskip} 
$\sqrt{e}\sin{\omega}$                                                    &  $-0.10 \pm 0.15$                         &      0 (fixed)            \\ \noalign {\smallskip} 
$c1_{V}$                                                               & $1.167 \pm 0.058$                    &   $1.168 \pm 0.058$                         \\ \noalign {\smallskip}  
$c2_{V}$                                                               & $0.017 \pm 0.056$                    & $0.016 \pm 0.061$              \\ \noalign {\smallskip}                             
$c1_{Gunn-r}$                                                    & $0.983 \pm 0.050$                     &  $0.985 \pm 0.050$                          \\ \noalign {\smallskip}  
$c2_{Gunn-r}$                                                    & $-0.121 \pm 0.044$                    &  $-0.122 \pm 0.045$                        \\ \noalign {\smallskip}                             
$c1_{BB}$                                                            & $1.040 \pm 0.045$                     &  $1.041 \pm 0.045$  \\ \noalign {\smallskip}                            
$c2_{BB}$                                                            & $-0.129 \pm 0.053$                    &  $-0.125 \pm 0.051$                           \\ \noalign {\smallskip}      
$c1_{I+z}$                                                            & $0.859 \pm 0.053$                     &   $0.856 \pm 0.053$                            \\ \noalign {\smallskip}                            
$c2_{I+z}$                                                            & $-0.206 \pm 0.041$                    &   $-0.208 \pm 0.0241$                              \\ \noalign {\smallskip}                            
$T_{eff}$ [K] $^a$                                                        & $5400 \pm 100$              &   $5400 \pm 100$       \\ \noalign {\smallskip} 
{[Fe/H]} $^a$                                                                 &$-$0.08 $\pm$ 0.11            &         $-$0.08 $\pm$ 0.11 \\\noalign {\smallskip} 
\hline \noalign {\smallskip}
$Deduced$ $stellar$ $parameters$   &  &  \\ \noalign {\smallskip}
\hline \noalign {\smallskip}
Density $\rho_* $  [$\rho_\odot $]                       & $0.90_{-0.09}^{+0.16}$ &  $0.849_{-0.044}^{+0.053}$       \\ \noalign {\smallskip} 
Surface gravity $\log g_*$ [cgs]                          & $4.406_{-0.029}^{+0.044}$      &  $4.392 \pm 0.016$   \\ \noalign {\smallskip} 
Mass $M_\ast $    [$M_\odot$]                            & $0.993_{-0.037}^{+0.034}$       &  $1.004 \pm 0.028$ \\ \noalign {\smallskip} 
Radius  $ R_\ast $   [$R_\odot$]                         & $1.036_{-0.065}^{+0.046}$       &  $1.058 \pm 0.025 $       \\ \noalign {\smallskip} 
Luminosity $L_\ast$ [$L_\odot$]                        & $0.90 \pm 0.15$             &    $0.95 \pm 0.13$          \\ \noalign {\smallskip} 
$u1_{V}$                                                              & $0.470 \pm 0.033$           &   $0.470 \pm 0.035$      \\ \noalign {\smallskip} 
$u2_{V}$                                                              & $0.226 \pm 0.027$           &    $0.227 \pm 0.028$      \\ \noalign {\smallskip} 
$u1_{Gunn-r}$                                                    & $0.369 \pm 0.027$           &    $0.370 \pm 0.027$        \\ \noalign {\smallskip} 
$u2_{Gunn-r}$                                                    & $0.245 \pm 0.020$           &   $0.246 \pm 0.020$      \\ \noalign {\smallskip} 
$u1_{BB}$                                                          & $0.391 \pm 0.025$            &  $0.392 \pm 0.025$        \\ \noalign {\smallskip} 
$u2_{BB}$                                                          & $0.259 \pm 0.023$            &  $0.259 \pm 0.022$      \\ \noalign {\smallskip} 
$u1_{I+z}$                                                          & $0.302 \pm 0.028$            &   $0.301 \pm 0.028$      \\ \noalign {\smallskip} 
$u2_{I+z}$                                                          & $0.255 \pm 0.017$            &   $0.255 \pm 0.018$   \\ \noalign {\smallskip} 
\hline \noalign {\smallskip}
$Deduced$ $planet$ $parameters$   & &   \\ \noalign {\smallskip}
\hline \noalign {\smallskip}
RV $K$ [\ms]                                                         & $214 \pm 13$           &   $221 \pm 11$                \\ \noalign {\smallskip} 
$R_p/R_\ast$                                                       & $0.1234 \pm 0.0011$    & $0.1234 \pm 0.0011$                 \\ \noalign {\smallskip} 
$b_{tr}$ [$R_\ast$]                                               & $0.313_{-0.064}^{+0.049}$ &  $0.322_{-0.068}^{+0.048}$          \\ \noalign {\smallskip} 
$b_{oc}$ [$R_\ast$]                                             & $0.296_{-0.070}^{+0.059}$    & $0.322_{-0.068}^{+0.048}$ \\ \noalign {\smallskip} 
$T_{oc}-2450000$ [$HJD_{TDB}$]                  &  $5583.403_{-0.024}^{+0.026}$            &   $5583.38834 \pm 0.00029$     \\ \noalign {\smallskip} 
Orbital semi-major axis $ a $ [AU]                    &  $0.02640_{-0.00033}^{+0.00030}$        &   $0.02648 \pm 0.00024$       \\ \noalign {\smallskip} 
$a / R_\ast$                                                            & $5.49_{-0.19}^{+0.31}$                & $5.39_{-0.09}^{+0.11}$  \\ \noalign {\smallskip} 
Orbital inclination $i_p$ [deg]                               & $86.69_{-0.66}^{+0.79}$              &  $86.57_{-0.60}^{+0.80}$ \\ \noalign {\smallskip} 
Orbital eccentricity $ e $                                       & $0.035_{-0.025}^{+0.039}$, $<0.132$ (95\%)           &     0 (fixed) \\ \noalign {\smallskip}
 Argument of periastron  $ \omega $ [deg]          & $308_{-36}^{+80}$                         & -   \\ \noalign {\smallskip} 
Equilibrium temperature $T_{eq}$ [K]$ $$^b$    & $1672_{-63}^{+59}$                     & $1689 \pm 49$  \\ \noalign  {\smallskip} 
Density  $ \rho_p$ [$\rho_{\rm Jup}$]                    &  $0.64_{-0.08}^{+0.12}$             &   $0.619_{-0.052}^{+0.064}$    \\ \noalign {\smallskip} 
Surface gravity $\log g_p$ [cgs]                             &  $3.294_{-0.043}^{+0.049}$                         &  $3.292 \pm 0.030$ \\ \noalign  {\smallskip} 
Mass  $ M_p$ [$M_{\rm Jup}$]                               & $1.217 \pm 0.083$                                        &    $1.271 \pm 0.068$  \\ \noalign {\smallskip} 
Radius  $ R_p $ [$R_{\rm Jup}$]                            & $1.244_{-0.080}^{+0.062}$                        & $1.271 \pm 0.039$  \\ \noalign {\smallskip} 
\hline \noalign {\smallskip}
\end{tabular}
\caption{Median and 1-$\sigma$ limits of the marginalized posterior distributions obtained for the WASP-64 
system parameters as derived from our MCMC analysis. $^a$Using as priors the spectroscopic values given in Table 3. $^b$Assuming a null Bond albedo ($A_B$=0) and isotropic reradiation ($f$=1/4). }
\end{center}
\end{table*}

\begin{table*}[h]
\begin{center}
\begin{tabular}{ccc}
& {\bf WASP-72}  & \\
\hline \noalign {\smallskip}
$Free$ $parameters$ &  $e \geq 0 $ & $e = 0$ {\bf (adopted)} \\ \noalign {\smallskip}
\hline \noalign {\smallskip}
Planet/star area ratio  $ (R_p/R_\ast)^2 $  [\%]         &  $0.423_{-0.037}^{+0.039}$               & $0.430_{-0.039}^{+0.043}$ \\ \noalign {\smallskip} 
$b'=a\cos{i_p}/R_\ast$ [$R_*$]                                    &  $0.58_{-0.20}^{+0.10}$                      &  $0.59_{-0.18}^{+0.10}$       \\ \noalign {\smallskip} 
Transit width  $W$ [d]                                                     &   $0.1552 \pm 0.0029$                  & $0.1558_{-0.0029}^{+0.0035}$  \\ \noalign {\smallskip} 
$T_0-2450000$ [$HJD_{TDB}$]                                 &  $5583.6525 \pm 0.0021$                   & $5583.6529 \pm 0.0021$       \\ \noalign {\smallskip}
Orbital period  $ P$ [d]                                       &  $2.2167434_{-0.0000077}^{+0.0000084}$    & $2.2167421 \pm 0.0000081$    \\ \noalign {\smallskip} 
RV $K_2$  [m\,s$^{-1}$\,d$^{1/3}$]                             & $236.6 \pm 5.6$                                    &  $236.1 \pm 5.5$                          \\ \noalign {\smallskip} 
RV $\gamma$ [km\,s$^{-1}$]                                        & $35.923 \pm 0.015$                             &  $35.919 \pm 0.014$   \\ \noalign {\smallskip} 
RV slope [m\,s$^{-1}$\,y$^{-1}$]                         &  $-44 \pm 19$                                                 &   $-39 \pm 17$                \\ \noalign {\smallskip} 
$\sqrt{e}\cos{\omega}$                                         &  $-0.022 \pm 0.071$                                      &  0 (fixed)                         \\ \noalign {\smallskip} 
$\sqrt{e}\sin{\omega}$                                          &  $-0.06 \pm 0.11$                                          &   0 (fixed)                        \\ \noalign {\smallskip} 
$c1_{Gunn-r}$                                                       & $0.938 \pm 0.026$                                        &  $0.936 \pm 0.026$          \\ \noalign {\smallskip}  
$c2_{Gunn-r}$                                                       & $-0.318 \pm 0.016$                                       & $-0.319 \pm 0.017$      \\ \noalign {\smallskip}                             
$c1_{BB}$                                                            & $0.820 \pm 0.082$                                          &  $0.818 \pm 0.078$  \\ \noalign {\smallskip}                            
$c2_{BB}$                                                            & $-0.352 \pm 0.046$                                         & $-0.352 \pm 0.042$   \\ \noalign {\smallskip}      
$c1_{I+z}$                                                            & $0.713 \pm 0.041$                                          &  $0.710 \pm 0.042$   \\ \noalign {\smallskip}                            
$c2_{I+z}$                                                            & $-0.382 \pm 0.028$                                         &  $-0.383 \pm 0.027$  \\ \noalign {\smallskip}         
$T_{eff}$ [K] $^a$                                                   & $6250 \pm 100$                                              &  $6250 \pm 100$                         \\ \noalign {\smallskip} 
[Fe/H] $^a$                                                                  &$-$0.06 $\pm$ 0.09                                        &  $-$0.06 $\pm$ 0.09                    \\ \noalign {\smallskip}                    
\hline \noalign {\smallskip}
$Deduced$ $stellar$ $parameters$   &  &  \\ \noalign {\smallskip}
\hline \noalign {\smallskip}
Surface gravity $\log g_*$ [cgs] $^a$                      & $3.99 \pm 0.10$                      &  $3.99_{-0.11}^{+0.10}$              \\ \noalign {\smallskip} 
Density $\rho_* $  [$\rho_\odot $]                       & $0.181_{-0.046}^{+0.074}$                      &  $0.177_{-0.048}^{+0.073}$       \\ \noalign {\smallskip} 
Mass $M_\ast $    [$M_\odot$]                            & $1.382 \pm 0.053$                       & $1.386 \pm 0.055$        \\ \noalign {\smallskip} 
Radius  $ R_\ast $   [$R_\odot$]                         & $1.97 \pm 0.23$                             & $1.98 \pm 0.24$                \\ \noalign {\smallskip} 
Luminosity $L_\ast$ [$L_\odot$]                        & $5.3_{-1.2}^{+1.4}$                                    &  $5.3_{-1.3}^{+1.5}$                      \\ \noalign {\smallskip} 
$u1_{Gunn-r}$                                                       & $0.311 \pm 0.013$                                     &  $0.311 \pm 0.013$                        \\ \noalign {\smallskip} 
$u2_{Gunn-r}$                                                    & $0.3148 \pm 0.0063$                                   &   $0.3147 \pm 0.0060$                  \\ \noalign {\smallskip} 
$u1_{BB}$                                                          & $0.257 \pm 0.043$                                         &  $0.256 \pm 0.041$                        \\ \noalign {\smallskip} 
$u2_{BB}$                                                          & $0.305 \pm 0.013$                                         &  $0.305 \pm 0.014$                        \\ \noalign {\smallskip} 
$u1_{I+z}$                                                          & $0.209 \pm 0.022$                                         &   $0.208 \pm 0.022$                        \\ \noalign {\smallskip} 
$u2_{I+z}$                                                          & $0.295 \pm 0.012$                                          &   $0.295 \pm 0.013$                       \\ \noalign {\smallskip} 
\hline \noalign {\smallskip}
$Deduced$ $planet$ $parameters$   & &   \\ \noalign {\smallskip}
\hline \noalign {\smallskip}
RV $K$ [\ms]                                                        & $181.3 \pm 4.2$                                               & $181.0 \pm 4.2$                \\ \noalign {\smallskip} 
$R_p/R_\ast$                                                       & $0.0650 \pm 0.0029$                    &  $0.0656 \pm 0.0031$  \\ \noalign {\smallskip} 
$b_{tr}$ [$R_\ast$]                                               & $0.57_{-0.20}^{+0.10}$                                &  $0.59_{-0.18}^{+0.10}$                \\ \noalign {\smallskip} 
$b_{oc}$ [$R_\ast$]                                             & $0.58_{-0.21}^{+0.11}$                                & $0.59_{-0.18}^{+0.10}$                 \\ \noalign {\smallskip} 
$T_{oc}-2450000$ [$HJD_{TDB}$]                &   $5584.758 \pm 0.042$                   & $5584.7612 \pm 0.0021$             \\ \noalign {\smallskip} 
Orbital semi-major axis $ a $ [AU]                       &  $0.03705 \pm 0.00047$    & $0.03708 \pm 0.00050$  \\\noalign{\smallskip} 
$a / R_\ast$                                                          & $4.05_{-0.38}^{+0.49}$                                   & $4.02_{-0.40}^{+0.49}$  \\ \noalign {\smallskip} 
Orbital inclination $i_p$ [deg]                               & $81.8_{-2.5}^{+3.5}$                                   &  $81.6_{-2.6}^{+3.2}$ \\ \noalign {\smallskip} 
Orbital eccentricity $ e $                                       & $0.014_{-0.010}^{+0.018}$, $<0.079$ (95\%) &   0 (fixed) \\ \noalign {\smallskip}
Argument of periastron  $ \omega $ [deg]            & $115_{-47}^{+111}$                                     &  -   \\ \noalign {\smallskip} 
Equilibrium temperature $T_{eq}$ [K]$ $$^b$      & $2200_{-120}^{+110}$                            &$2210_{-130}^{+120}$  \\ \noalign  {\smallskip} 
Density  $ \rho_p$ [$\rho_{\rm Jup}$]                         &  $0.75_{-0.25}^{+0.45}$                       & $0.72_{-0.25}^{+0.43}$    \\ \noalign {\smallskip} 
Surface gravity $\log g_p$ [cgs]                           &  $3.37_{-0.11}^{+0.13}$                              & $3.36 \pm 0.12$ \\ \noalign  {\smallskip} 
Mass  $ M_p$ [$M_{\rm Jup}$]                                   & $1.459 \pm 0.056$                  &  $1.5461_{-0.056}^{+0.059}$    \\ \noalign {\smallskip} 
Radius  $ R_p $ [$R_{\rm Jup}$]                                &    $1.25 \pm 0.19$                     &  $1.27 \pm 0.20$   \\ \noalign {\smallskip} 
\hline \noalign {\smallskip}
\end{tabular}
\caption{ Median and 1-$\sigma$ limits of the marginalized posterior distributions obtained for the WASP-72
system parameters as derived from our MCMC analysis. $^a$Using as priors the spectroscopic values given in Table 3. $^b$Assuming a null Bond albedo ($A_B$=0) and isotropic reradiation ($f$=1/4). }
\end{center}
\end{table*}

\begin{table}
\begin{center}
\begin{tabular}{cccc}
\hline \noalign {\smallskip}
Planet & Epoch &  Transit timing - 2450000 & $O-C$  \\ \noalign {\smallskip}
& & [$HJD_{TDB}$] & [$s$] \\ \noalign {\smallskip}
\hline \noalign {\smallskip}
WASP-64\,b & 0   &  $5582.60070 \pm 0.00051$     & $-85 \pm 50$  \\ \noalign {\smallskip} 
& 21  & $5615.64057 \pm 0.00057$     & $-19 \pm 55$  \\ \noalign {\smallskip} 
& 47  & $5656.54655  \pm 0.00035$    & $+15 \pm 39$ \\ \noalign {\smallskip} 
& 173 &  $5854.78224 \pm 0.00027$   & $+97 \pm 40$ \\ \noalign {\smallskip} 
& 173 &  $5854.78091 \pm 0.00032$   & $-19 \pm 41$  \\ \noalign {\smallskip} 
& 180  &  $5865.79456 \pm 0.00025$  & $+35 \pm 39$ \\ \noalign {\smallskip} 
& 194  &  $5887.82045 \pm 0.00035$   & $+18 \pm 44$ \\ \noalign {\smallskip} 
& 199  &  $5895.68667 \pm 0.00029$   & $-5 \pm 42$    \\ \noalign {\smallskip} 
& 206 &  $5906.69953 \pm 0.00041$    & $-19 \pm 50$  \\ \noalign {\smallskip} 
& 213  &  $5917.71279 \pm 0.00019$   & $-1 \pm 38$    \\ \noalign {\smallskip} 
& 229  &  $5939.73781 \pm 0.00030$  & $-92 \pm 45$   \\[0.33cm]
WASP-72\,b & 0   &  $5583.6542 \pm 0.0024$     & $+3.3 \pm 5.0$  \\ \noalign {\smallskip} 
& 125  & $5860.7441 \pm 0.0048$     & $-0.7\pm 7.7$  \\ \noalign {\smallskip} 
& 139  & $5891.7797  \pm 0.0017$    & $+0.9 \pm 4.6$ \\ \noalign {\smallskip} 
& 143 &  $5900.6475 \pm 0.0043$   & $+3.4 \pm 8.5$ \\ \noalign {\smallskip} 
& 300 &  $6248.6710 \pm 0.0029$   & $-5.0 \pm 6.4$ \\ \noalign {\smallskip} 
& 300 &  $6248.6758 \pm 0.0012$   & $+1.8 \pm 5.3$ \\ \noalign {\smallskip} 
& 309 &  $6268.6292 \pm 0.0024$   & $+5.7 \pm 6.3$ \\ \noalign {\smallskip} 
\hline \noalign {\smallskip}
\end{tabular}
\caption{Transit timings and TTVs  ($O-C$ = observed minus computed timing) 
derived from the MCMC global analyzes of the WASP-64\,b and  WASP-72\,b transits. }
\end{center}
\end{table}

\section{Discussion}

WASP-64\,b and WASP-72\,b are thus two new very short-period (1.57d and 2.22d)
planets slightly more massive than Jupiter orbiting moderately bright ($V$=12.3 and 10.1) 
Southern stars. Their detection demonstrates nicely  the high photometric potential
 of the WASP transit survey  (Pollacco et al. 2006), as both planets show transits of very low-amplitude 
($<$ 0.5\%) in the WASP data. For WASP-64, the reason is not the intrinsic size contrast with the 
 host star but the dilution of the signal by a close-by spectroscopic binary that could be dynamically 
 bound to WASP-64 (Sec. 3.1). 
 
Fig.~13 shows the position of both planets in mass-radius  and irradiation-radius diagrams 
for a mass range from 1 to 2 $M_{Jup}$. WASP-64\,b  lies in a well-populated
area of  the irradiation-radius diagram. Its physical dimensions can be considered as rather standard. 
Its measured radius of  $1.27 \pm 0.04$ $R_{\rm Jup}$ agrees well with the value of $1.22 \pm 
0.11$ $R_{Jup}$ predicted by the equation derived by Enoch et al. (2012) from a sample of 71 
transiting planets with a mass between 0.5 and 2  $M_{\rm Jup}$ and relating planets' sizes to
their equilibrium temperatures and semi-major axes. On the contrary, WASP-72\,b 
 appears to be a possible outlier,  its measured radius of $1.27 \pm 0.20$ $R_{\rm Jup}$ 
 being marginally lower than the value predicted by Enoch et al's empirical relation, 
 $1.70 \pm 0.11$ $R_{Jup}$. Its density of $0.732_{-0.25}^{+0.43}$ $\rho_{\rm Jup}$ 
 could indeed be considered as  surprisingly high given its  extreme irradiation 
 ($\sim5.5\times10^9$ erg\,s$^{-1}$\,cm$^{-2}$), suggesting a  possible enrichment of heavy elements. 
 Nevertheless, Fig.~13 clearly shows that the errors
on its physical parameters  are still too high to draw any strong inference on 
its internal structure or its possible peculiarity.  Indeed, the transit parameters are still loosely determined, 
especially the impact parameter, resulting in a high relative error on the  stellar density that propagates to the
 stellar and planetary radii. Planets with similar irradiation and mass being still rare, it will thus be especially interesting
  to improve these parameters with new follow-up observations. 
  
 As described in Sec. 3.2, we have adopted the circular orbital models for both 
 planets, basing on their higher marginal likelihoods. To assess the validity of this choice
 on a theoretical basis, we integrated the future orbital  evolution of both systems through 
 the low-eccentricity tidal model presented by Jackson et al. (2008), assuming as starting eccentricity 
 the 95\% upper limits derived in our MCMC analysis (0.132 for WASP-64\,b and 0.052 for WASP-72\,b) 
 and assuming mean values of $10^{7}$ and $10^{5}$ for, respectively, the stellar and planetary tidal
  dissipation parameters $Q'_\ast$ and  $Q'_p$ (Goldreich \& Soter, 1966). These integrations resulted 
  in circularization times (defined as the time needed to reach $e<0.001$) of 4 and  24 Myr for, respectively, 
  WASP-64\,b and WASP-72\,b. These times are much shorter than the estimated age of the systems, strengthening
our selection of the circular orbital solutions. It is worth mentioning that both systems are tidally unstable, 
as most of the hot Jupiter systems (Levrard et al. 2009),  the times remaining for the planets to reach their Roche limits 
being respectively 0.9 Gy (WASP-64\,b) and  0.35 Gy (WASP-72\,b) under the assumed tidal dissipation parameters.

The new transiting systems reported here represent both two interesting targets for follow-up 
observations. Thanks to its extreme irradiation and its moderately high planet-to-star size ratio, 
WASP-64\,b is a good target for near-infrared occultation (spectro-)photometry programs able
to probe its day-side  spectral energy distribution. Assuming a null albedo for the planet
and blackbody spectra for both the planet and the host star, we computed occultation depths
of 650-1550 ppm in $K$-band, 1500-2750 ppm at 3.6 $\mu$m and 2050-3350 ppm at 4.5 $\mu$m, the 
lower and upper limits corresponding, respectively, to a uniform redistribution of the stellar radiation
 to both planetary hemispheres and to a direct reemission of the dayside hemisphere. Precise
measurement of such occultation depths is definitely within the reach of several ground-based and
 space-based facilities (e.g. Gillon et al. 2012). For ground-based programs, the situation is made easier
  by the presence of a bright ($K=9.8$) comparison star at only 28'' from the target (Fig.~3). For
  WASP-72, atmospheric measurements are certainly more challenging considering the lower
  planet-to-star size ratio. Here, the first follow-up actions should certainly be to  confirm and improve
  our measurement of the planet's size through high-precision transit photometry, and to gather more RVs  
  to confirm the trend marginally detected in our analysis.
\begin{figure*}
\label{fig:mr}
\centering                     
\includegraphics[width=8.5cm]{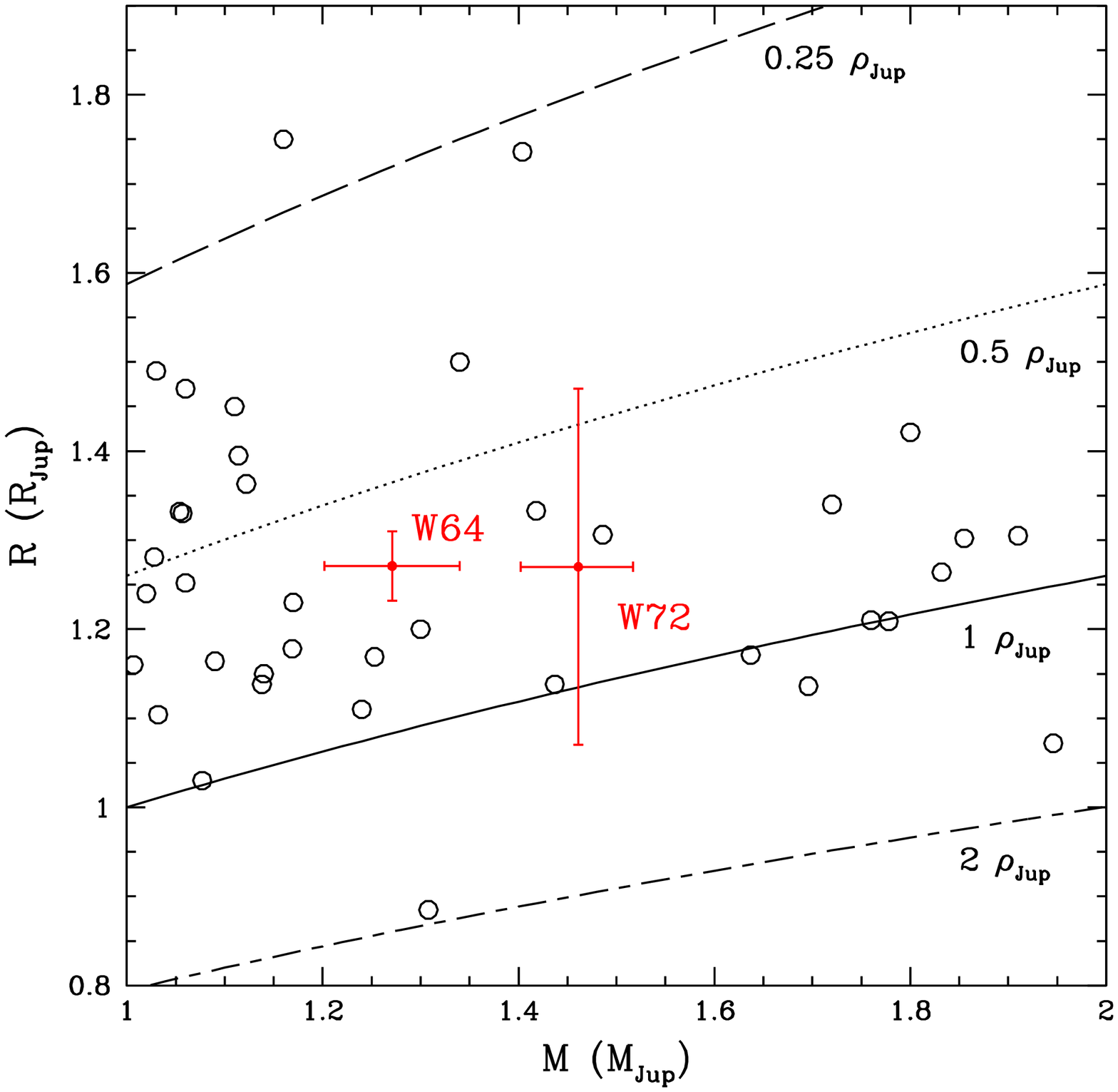}
\includegraphics[width=8.5cm]{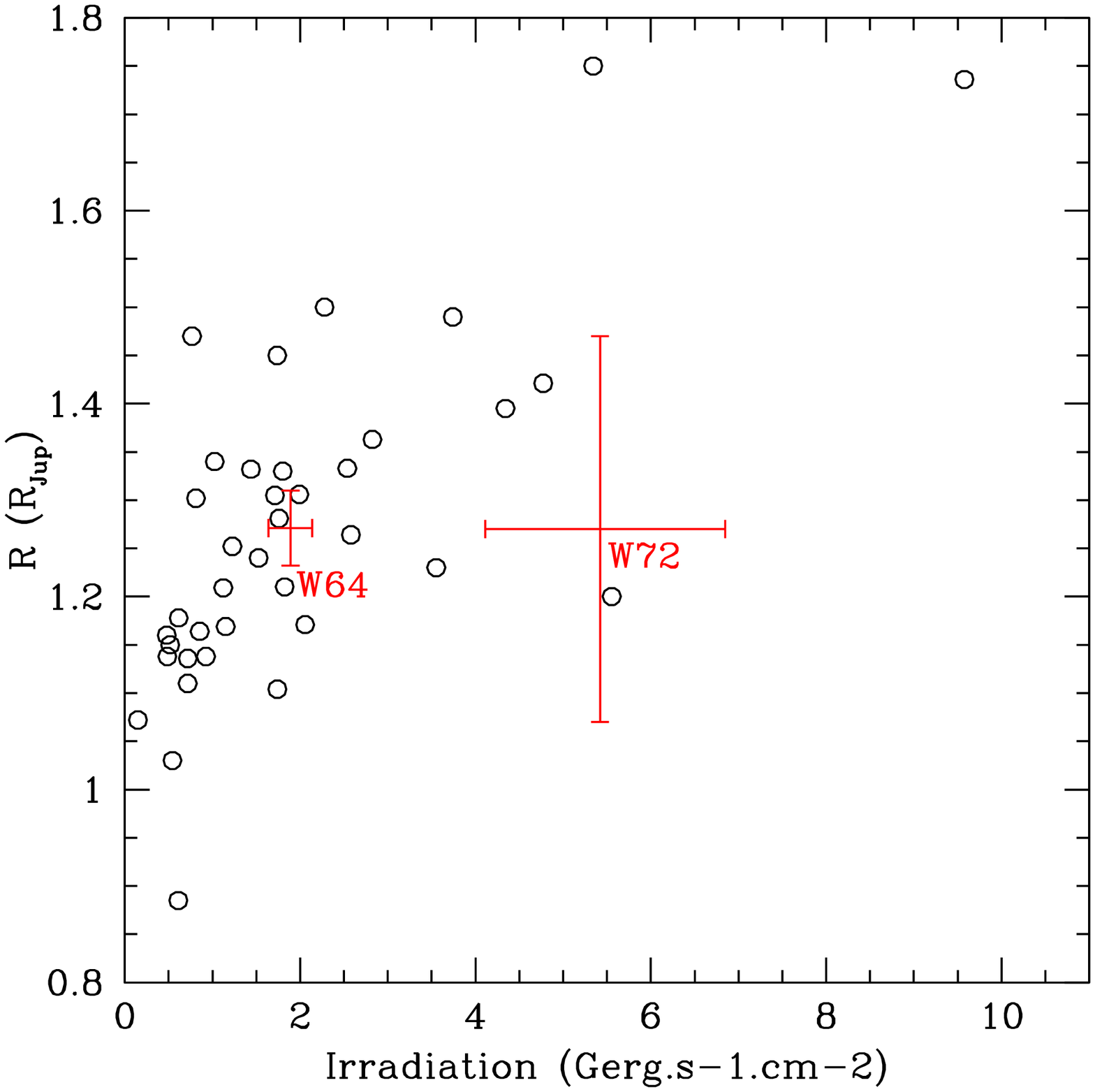}
\caption{$Left$: mass--radius diagram for the transiting planets with masses
ranging from 1 to 2 $M_{\rm Jup}$ (data from exoplanet.eu, Schneider et al. 2011). 
WASP-64\,b and WASP-72\,b are shown as red square symbols, while the other planets are 
shown as open circles without their error bars for the sake of clarity. Different iso-density lines are also 
shown. $Right$: position of WASP-64\,b and WASP-72\,b in a
irradiation--radius diagram for the same exoplanets sample.}
\end{figure*}

\begin{acknowledgements}
WASP-South is hosted by the South African Astronomical Observatory and we are grateful for their ongoing support and assistance. Funding for WASP comes from consortium universities and from UK's Science and Technology Facilities Council. TRAPPIST is a project funded by the Belgian Fund for Scientific Research (Fond National de la Recherche Scientifique, F.R.S-FNRS) under grant FRFC 2.5.594.09.F, with the participation of the Swiss National Science Fundation (SNF).  M. Gillon and E. Jehin are FNRS Research Associates. 
We thank the anonymous referee for his useful criticisms and suggestions that helped us to improve the
quality of the present paper.
\end{acknowledgements} 

\bibliographystyle{aa}

\end{document}